\documentclass[iop]{emulateapj}
\newcommand*{\teff}{$T_{\rm eff}$}
\newcommand*{\logg}{$\log~g$}
\newcommand*{\feh}{[Fe/H]}
\newcommand*{\afe}{[$\alpha$/Fe]}
\newcommand*{\kms}{km s$^{-1}$}

\newcommand*{\vphi}{$V_{\rm \phi}$}

\newcommand*{\alp}{$\alpha$}
\newcommand*{\rsun}{$R_\odot$}

\newcommand*{\z}{$|Z|$}
\newcommand*{\kmskpc}{km s$^{-1}$~kpc$^{-1}$}
\newcommand*{\kmsdex}{km s$^{-1}$~dex$^{-1}$}
\newcommand*{\dexkpc}{dex~kpc$^{-1}$}

\usepackage{epsfig}
\usepackage{apjfonts}
\usepackage{amsmath}
\usepackage{amssymb}
\usepackage{longtable}
\usepackage{leqno}
\usepackage{graphics}
\shorttitle{Formation and Evolution of the Milky Way Disk System}
\shortauthors{Lee et al.}
\slugcomment{Accepted for publication in The Astrophysical Journal on June 28, 2011}

\begin{document}

\title{Formation and Evolution of the Disk System of the Milky Way: \\
       \afe\ Ratios and Kinematics of the SEGUE G-Dwarf Sample}
\author{Young Sun Lee\altaffilmark{1}, Timothy C. Beers\altaffilmark{1},
Deokkeun An\altaffilmark{2}, \v Zeljko Ivezi\'c\altaffilmark{3},
Andreas Just\altaffilmark{4}, \\ Constance M. Rockosi\altaffilmark{5},
Heather L. Morrison\altaffilmark{6}, Jennifer A. Johnson\altaffilmark{7},
Ralph Sch\"onrich\altaffilmark{8}, \\ Jonathan Bird\altaffilmark{7},
Brian Yanny\altaffilmark{9}, Paul Harding\altaffilmark{6}, and Helio J. Rocha-Pinto\altaffilmark{10,11}}

\altaffiltext{1}{Department of Physics \& Astronomy and JINA (Joint Institute for Nuclear Astrophysics),
                 Michigan State University, East Lansing, MI 48824, USA; lee@pa.msu.edu, beers@pa.msu.edu}
\altaffiltext{2}{Department of Science Education, Ewha Womans University, Seoul 120-750, Republic of Korea}
\altaffiltext{3}{Astronomy Department, University of Washington, Box 351580, Seattle, WA 98195-1580, USA}
\altaffiltext{4}{Astronomisches Rechen-Institut, Zentrum f\"ur Astronomie der
                 Universit\"at Heidelberg (ZAH), M\"onchhofstr. 12-14, 69120 Heidelberg, Germany}
\altaffiltext{5}{UCO/Lick Observatory, Department of Astronomy and Astrophysics,
                 University of California, Santa Cruz, CA 95064, USA}
\altaffiltext{6}{Department of Astronomy, Case Western Reserve University, Cleveland, OH 44106, USA}
\altaffiltext{7}{Department of Astronomy, Ohio State University, Columbus, OH 43210, USA}
\altaffiltext{8}{Max-Planck-Institute f\"ur Astrophysik, Karl-Schwarzschild-Str. 1, D-85741, Garching, Germany}
\altaffiltext{9}{Fermi National Accelerator Laboratory, Batavia, IL 60510, USA}
\altaffiltext{10}{Universidade Federal do Rio de Janeiro, Observat\'orio do Valongo,
                  Lad. Pedro Ant\^onio 43, 20080-090 Rio de Janeiro, Brazil}
\altaffiltext{11}{Laborat\'orio Interinstitucional de e-Astronomia - LIneA, Rua Gal.
                  Jos\'e Cristino 77, 20921-400 Rio de Janeiro, Brazil}

\begin{abstract}
We employ measurements of the \afe\ ratio derived from low-resolution
($R\sim2000$) spectra of 17,277 G-type dwarfs from the SEGUE survey to separate
them into likely thin- and thick-disk subsamples. Both subsamples exhibit strong
gradients of orbital rotational velocity with metallicity, of opposite signs,
$-$20 to $-$30 \kmsdex\ for the thin-disk and $+$40 to $+$50 \kmsdex\ for the
thick-disk population. The rotational velocity is uncorrelated with
Galactocentric distance for the thin-disk subsample, and exhibits a small trend
for the thick-disk subsample. The rotational velocity decreases with distance
from the plane for both disk components, with similar slopes ($-9.0 \pm 1.0$
\kmskpc). Thick-disk stars exhibit a strong trend of orbital eccentricity with
metallicity (about $-0.2~\rm dex^{-1}$), while the eccentricity does not change
with metallicity for the thin-disk subsample. The eccentricity is almost
independent of Galactocentric radius for the thin-disk population, while a
marginal gradient of the eccentricity with radius exists for the thick-disk
population. Both subsamples possess similar positive gradients of eccentricity
with distance from the Galactic plane. The shapes of the eccentricity
distributions for the thin- and thick-disk populations are independent of
distance from the plane, and include no significant numbers of stars with
eccentricity above 0.6. Among several contemporary models of disk evolution we
consider, radial migration appears to have played an important role in the
evolution of the thin-disk population, but possibly less so for the thick disk,
relative to the gas-rich merger or disk heating scenarios. We emphasize that
more physically realistic models and simulations need to be constructed in order
to carry out the detailed quantitative comparisons that our new data enable.
\end{abstract}


\section{Introduction}

The Milky Way's thick disk, first identified from fits of the vertical density
profile of stars with a mix of exponential functions (Yoshii 1982; Gilmore \&
Reid 1983), differs in many ways from the thin disk, e.g., in its
kinematics and chemical abundances.

The scale height of the thick disk is about 1 kpc, while that of the thin disk
is $\sim0.3$ kpc. Typical thick-disk stars have generally lower net
orbital rotational velocities with larger velocity dispersions (Majewski 1993;
Chiba \& Beers 2000; Robin et al. 2003; Soubiran et al. 2003; Parker et al.
2004; Wyse et al. 2006), possess higher \afe\ ratios\footnote[12]{The \afe\ ratio is
often represented by an average of the [Mg/Fe], [Si/Fe], [Ca/Fe], and [Ti/Fe]
ratios, which we adopt in this paper as well.}, and are older and more
metal-poor than typical thin-disk stars (Bensby et al. 2003, 2005; Reddy et al.
2006, 2010; Fuhrmann 2008; Haywood 2008).

Their higher \afe\ ratios and older ages imply that thick-disk stars were
born earlier than most thin-disk stars, in an environment of rapid star
formation, and that they have likely had more time to experience dynamical heating
and secular processes such as scattering by perturbations in the disk. As a
result of the multiple complex processes that thick-disk stars may have
experienced during their lifetimes, consensus on the nature of the formation and
evolution of the thick disk has yet to be reached.

The currently discussed mechanisms for thick-disk formation can be broadly
divided into two groups -- violent origin and secular evolution. Among the
models involving violent origin, the heating scenario (e.g., Quinn et al. 1993;
Kazantzidis et al. 2008) posits that the thick disk results from a pre-existing
thin disk that has been dynamically heated by satellite mergers. In their
simulations of this process, Villalobos \& Helmi (2008) found that on the
order of 10--20\% of the stars in the thickened disk component were accreted
from satellites, the rest being heated thin-disk stars. The accretion origin
of the thick disk (e.g., Abadi et al. 2003) invokes the hypothesis that
thick-disk stars were predominantly formed in dwarf-like galaxies, which were
then directly assimilated into the thick disk from orbits that reached near the
Galactic disk plane. Abadi et al. (2003) predicted that over 70\% of thick-disk
stars were accreted from such disrupted galaxies. The third model among the
violent origin class is that thick-disk stars may have formed $in~situ$ through
chaotic mergers of gas-rich systems, prompting simultaneous early star formation
before and during the mergers (Brook et al. 2004, 2005, 2007), and that
thin-disk stars formed after the merger events settled down.

Secular evolution by disk heating was first conceived by Spitzer \&
Schwarzschild (1953), who demonstrated that encounters with molecular clouds
could increase the velocity dispersion of late type, old stars. Barbanis \&
Woltjer (1967) also showed that spiral structures might be the cause of the
larger velocity dispersion of older stars in the solar neighborhood. These ideas
had been further developed by several studies (e.g., Fuchs 2001, and references
therein). Although challenged by Jenkins (1992), disk heating by secular
processes have recently regained attention, both observationally and
theoretically, as possible thick-disk formation scenarios.

Indeed, recent theoretical studies and simulations (Sch\"onrich \& Binney 2009a,
2009b; Loebman et al. 2010) suggested that the thick disk might not require
a violent origin, but rather could have formed by cumulative secular
processes associated with the radial migration of stars. According to the
migration theories (Sellwood \& Binney 2002; Ro\v skar et al. 2008a), stars in
the Galactic disk can radially move from the inner (outer) to the outer (inner)
regions due to resonant scattering by transient spiral structure. Based on their
simulations, Minchev \& Famaey (2010) also suggested that long-lived spiral
structures, interacting with a central bar, could be responsible for the radial
movements of stars in a disk galaxy.

These proposed models predict various trends between the kinematic parameters and
chemical abundances of disk-system stars, as well as between their kinematics
and spatial distributions. For example, Sch\"onrich \& Binney (2008b) suggested
that local, relatively metal-rich thin-disk stars, formed in the inner part of
the disk and moved outward, while local metal-poor thin-disk stars were born in
the outer disk and migrated inward to the solar radius, retaining
information on the kinematic differences between the two populations. Thus,
there should exist a gradient in the variation of rotational velocity with
metallicity; evidence for such a behavior has been claimed observationally by
Haywood (2008). Models of disk heating via satellite mergers (Villalobos et al.
2010) result in proposed relationships between rotational velocity and
Galactocentric distance and distance from the Galactic plane. Gas-rich merger
models (Brook et al. 2007) also predict a gradient of rotational velocity with
Galactocentric radius for disk stars near the solar radius.

Sales et al. (2009) proposed that the distribution of orbital eccentricities for
nearby thick-disk stars could be used to provide constraints on the various
suggested formation models. A number of recent papers also have employed this
framework to study possible origins of the thick disk, based on data from
several large spectroscopic surveys. For example, Wilson et al. (2011) have
explored data from the RAdial Velocity Experiment (RAVE; Steinmetz et al. 2006),
while Dierickx et al. (2010) used data from the seventh public release of the
Sloan Digital Sky Survey (SDSS DR7; York et al. 2000; Abazajian et al. 2009).
The study of Casetti-Dinescu et al. (2011) combined RAVE data with newly
available proper motions from the fourth release of the Southern Proper Motion
Catalog (SPM4; Girard et al. 2011). We discuss their analyses and
conclusions further below.

Most previous observational studies that have sought to test the various
correlations predicted by the models mentioned above have used methods of
assigning individual stars to membership in the thin- and thick-disk populations
(based on a given star's location or kinematics) that introduce manifest biases
that can confound interpretations (as previously noted by Sch\"onrich \& Binney
2009b and Loebman et al. 2010). As the chemical signatures of a star are
substantially less variable properties than its spatial position or velocities
over its lifetime, it is instead desirable to classify disk stars into their
likely components according to their chemistry.

Among the various chemical abundance ratios that might be explored for this
purpose, the \afe\ ratios appear particularly useful. These ratios can be
relatively easily measured (as described below), and have been proven to well
separate thick-disk stars from thin-disk stars. It is known that, at least in
the solar neighborhood (where essentially all previous studies have been
conducted), thick-disk stars are on average enhanced in their \afe\ ratios by
$+0.2$ to $+0.3$ dex relative to their thin-disk counterparts at a given [Fe/H].
Local kinematically selected thin- and thick-disk samples based on probabilistic
membership assignments have confirmed this enhancement of \afe\ (e.g., Bensby et
al. 2003; Reddy et al. 2006). Fuhrmann (1998, 2008) also demonstrated that
dwarfs in his volume-limited sample could be clearly separated into two
populations as a function of [Fe/H] -- one associated with high [Mg/Fe] and the
other with low [Mg/Fe]. The elemental abundance patterns of the stars with low
[Mg/Fe] ratios and high [Mg/Fe] ratios are very similar to the kinematically
selected thin- and thick-disk samples. The recent study by Nissen \& Schuster
(2010) demonstrated that nearby dwarfs with halo kinematics could be separated
into two groups on \afe. They proposed that the high-\alp\ stars may have been
born in the disk or bulge of the milky way and heated to halo kinematics by
merging satellite galaxies or else were simply members of the early generations
of halo stars born during the collapse of a proto-Galactic gas cloud, while the
low-\alp\ stars may have been accreted from dwarf galaxies. Clearly, \afe\ 
ratios also provide valuable information on the timescales and intensities of
star formation in the populations involved.

In this study we make use of the first set of \afe\ ratios obtained for a large
sample of low-resolution ($R\sim2000$) spectra from the Sloan Extension for
Galactic Understanding and Exploration (SEGUE; Yanny et al. 2009). As shown by
Lee et al. (2011), for stars with SDSS/SEGUE spectra of signal-to-noise (S/N)
ratios greater than 20 \AA$^{-1}$, and with temperatures in the range 4500~K $\le$ \teff\
$\le$ 7000~K, one can estimate \afe\ with an accuracy of better than 0.1 dex
(as derived by comparing \afe\ estimates from the analysis of moderately
high-resolution and medium-resolution spectra with those obtained from
application of our methods to the same stars). This enables a
$chemical~separation$ of the disk system into likely thin- and thick-disk
populations. In this paper we explore the observed correlations of rotational
velocity and orbital eccentricity with metallicity, Galactocentric distance and
distance from the Galactic plane, as well as the orbital eccentricity
distributions for the individual populations, and compare with the predictions
of the radial migration, gas-rich merger, accretion, and dynamical heating
models. Since we believe that direct quantitative comparisons with the
predictions made by various models (or simulations) mentioned in this study are
somewhat premature, we emphasize the more qualitative aspects of these
comparisons.

This paper is outlined as follows. In Section 2 we present the G-dwarf sample
from SEGUE, describe various cuts imposed on the sample to obtain a refined disk
dwarf sample, and discuss the calculations used to derive their space motions
and orbital eccentricities. Section 3 describes how we assign membership of the
stars into either the thin- or thick-disk populations. Results from our G-dwarf sample and
discussion of comparisons of our results with the predictions of various contemporary
disk formation and evolution scenarios follow in Sections 4 and 5, respectively.
A summary and our conclusions follow in Section 6.

\section{Selection of Local Dwarf Stars}

\subsection{The SEGUE G-dwarf Sample}

Our initial sample comprises low-resolution ($R\sim$2000) spectra of
$\sim$63,000 stars from SDSS Data Release 8 (DR8; Aihara et al. 2011), obtained
during the SEGUE sub-survey, which were originally targeted as G-dwarf candidates
(with colors and magnitudes in the range 0.48 $< (g-r)_{0} <$ 0.55 and $r_{0} <$
20.2). As a result of the simple sampling function, this dataset is expected to
be relatively unbiased with respect to chemistry, and completely unbiased with
respect to kinematics. In order to obtain a subsample of disk stars with the
most reliably estimated physical quantities we apply several additional cuts.

First, we exclude stars lacking information on their stellar parameters
(effective temperature, \teff, surface gravity, \logg, and metallicity, \feh),
radial velocities, or proper motions. The stellar atmospheric parameters were
determined by the most recent version of the
SEGUE Stellar Parameter Pipeline (SSPP; Lee et al. 2008a,
2008b; Allende Prieto et al. 2008; Smolinski et al. 2011); typical external
errors in these estimates are 180 K in \teff, 0.24 dex in log $g$, and 0.23 dex
in [Fe/H] (Smolinski et al. 2011). It has been shown that shifts in the
SSPP-derived estimates of [Fe/H] and \afe\ caused by the presence of
unrecognized spectroscopic binaries are generally small (Schlesinger et al.
2010). Although the typical uncertainty of the radial velocity varies with the
S/N ratio of a spectrum, it is less than 5 \kms\ for the great majority of stars
in our sample. Proper motion information was obtained based on the procedures
described by Munn et al. (2004); the systematic error noted by Munn et al.
(2008) has been corrected (final typical errors are 3--4 mas yr$^{-1}$), and we
have adopted the Munn et al. recommendations for maximum fit residuals and minimum
numbers of epochs considered in order to obtain the most reliable proper
motions. In this regard, see also Bond et al. (2010), who investigated the
systematic errors in Munn et al. (2008) by comparison with the expected null
proper motions of SDSS quasars.

\begin{figure*}
\centering
\plotone{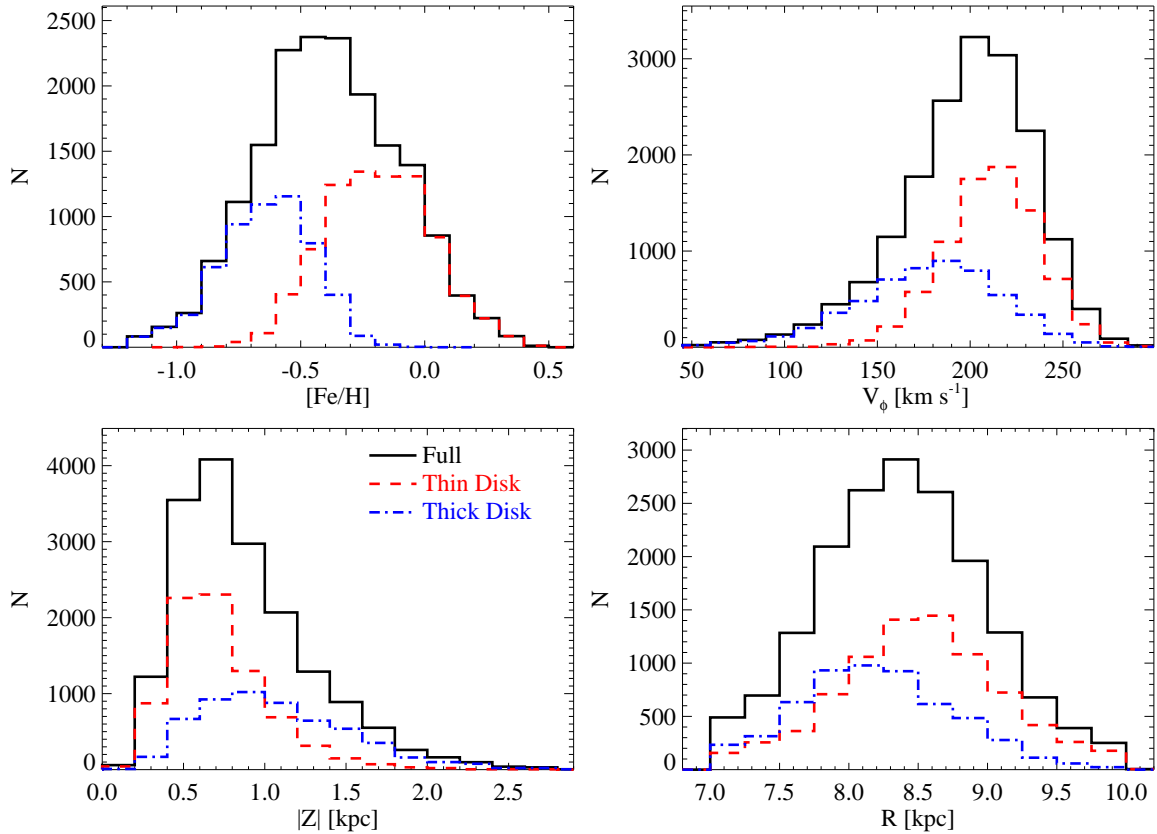}
%\plotone{sample_plot.eps}
\caption{Distributions of metallicities, [Fe/H] (top left), rotational
velocities, \vphi\ (top right), distances from the Galactic plane, \z\ (bottom
left), and Galactocentric distances projected onto the plane, $R$ (bottom right)
for the final selected sample. The solid line indicates our final full dwarf
sample, while the dashed red and dotted-dash blue lines are the thin- and
thick-disk subsamples, respectively, with stars assigned by the procedures
described in Section 3.}
\label{fig:sample}
\end{figure*}
%\newpage

Distances to individual stars are estimated using a calibrated set of stellar
isochrones (An et al. 2009a), following the prescription in An et al. (2009b).
After correcting photometry for dust extinction, main-sequence fitting is 
performed simultaneously on three different color-magnitude diagrams (CMDs),
with $r$ as a luminosity index, and $g-r$, $g-i$, and $g-z$ as color indices,
respectively. We adopt an SSPP-derived [Fe/H] in the distance estimation for
each star, and fix a stellar age and \afe\ of the model at a given [Fe/H],
assuming a linear relationship between [Fe/H] and these quantities (see An et
al. 2009b). Distance estimates obtain using \afe\ from this assumption may not be
internally consistent with analyses based on the SSPP-determined \afe,
but even a $\sim$0.1 dex difference in \afe\ has a negligible impact on the
derived distances ($\sim$0.01 mag in distance modulus). We also limit models
in the fitting to \logg\ $\geq 4.2$ to minimize possible distance bias from
stellar age effects near the main-sequence turn-off. An inter-comparison of
results from the three CMDs suggests that the internal error in the distance
modulus is $\sim$0.1 mag; an additional $\sim$0.1 mag error is expected from
the combined errors in age, [Fe/H], \afe, and $E(B-V)$. This suggests that the
associated distance-modulus error is $\sim$0.14 mag for individual stars. The
effects of binarity are more difficult to quantify, and are not included in this
error estimate (see An et al. 2007; Sesar et al. 2008).

The \afe\ ratio is derived following the procedures described by Lee et al.
(2011). Briefly summarizing, Lee et al. first generated a grid of synthetic
spectra, covering 4000~K $\leq T_{\rm eff} \leq$ 8000~K in steps of 250~K, 0.0
$\leq \log~g \leq $ 5.0 in steps of 0.2 dex, $-4.0 \leq \rm [Fe/H] \leq +0.4$ in
steps of 0.2 dex, and $-0.1 \leq [\alpha/{\rm Fe}] \leq +0.6$, in steps of 0.1
dex, then determined \afe\ by searching the grid for a synthetic spectrum that
best matches a given SDSS/SEGUE spectrum (in regions that are most influenced by
\afe). By comparing with a set of moderately high-resolution ($R = 15,000$) and
medium-resolution ($R = 6000$) spectra of SDSS/SEGUE stars, they demonstrated
the ability to measure \afe\ from SDSS/SEGUE spectra (with S/N $> 20$ \AA$^{-1}$) with
uncertainties less than 0.1 dex, for stars with atmospheric parameters in the
range \teff\ = [4500, 7000] K, \logg\ = [1.5, 5.0], and \feh\ =~[$-$1.4,
$+$0.3], over the full range of \afe\ considered. For stars with [Fe/H] $<
-1.4$, slightly higher S/N was required to achieve this precision (S/N $> 25$ \AA$^{-1}$).

In order to assemble a local dwarf sample, we only include stars with distances,
$d$, less than 3 kpc from the Sun, and with \logg\ $\geq 4.2$. These cuts ensure
that we are selecting likely dwarfs from which we can obtain accurate space
motions (i.e., that do not suffer from severe degradation due to propagation of
proper motion errors at larger distances). In order to perform a confident
separation of the thin- and thick-disk populations on the basis of \afe, we
further require that the spectra of the dwarf stars included in our analysis
have S/N $\geq 30$ \AA$^{-1}$. This conservative cut on S/N ensures not only
high quality estimates of \feh\ and \afe, but also that our program stars have
small errors in estimated radial velocity (less than 5 \kms).

\subsection{Calculations of Space Motions and Orbital Eccentricity}

With information on the distances, radial velocities, and proper motions for our
program stars in hand, we then derive the $U$, $V$, $W$ space velocity
components. We apply $(U,V,W)_\odot$ = ($11.10,12.24,7.25$) \kms\ (Sch\"onrich et
al. 2010) to adjust for the solar peculiar motions with respect to the Local
Standard of Rest (LSR). For the purpose of our analysis, we also make use of the
rotational velocity around the Galactic center in a cylindrical coordinate
system, \vphi, calculated assuming \rsun\ = 8.0 kpc and $V_{\rm LSR}$ = 220 \kms.
The Galactocentric distance projected onto the Galactic plane, $R$, and the vertical
distance from the Galactic plane, \z, are also obtained. In addition, by
adoption of an analytic St\"ackel-type gravitational potential (which includes a
flattened, oblate disk and a spherically-shaped massive dark halo; see Chiba \&
Beers 2000), we compute $r_{\rm apo}$ ($r_{\rm peri}$), the maximum (minimum)
distance from the Galactic center that a star reaches during its orbit, as well
as the orbital eccentricity, $e$, defined as ($r_{\rm apo}-$$r_{\rm peri}$)
/($r_{\rm apo}+r_{\rm peri}$). Errors in the derived kinematics and orbital
parameters for each star due to propagation of the errors in the observed
quantities (mostly dominated by distance and proper motion errors) are
determined by 1000 realizations of a Monte Carlo simulation.

We next remove stars from our sample with derived rotational velocities relative
to the Galactic center less than \vphi\ = $+$50 \kms, with [Fe/H] $\leq -$1.2,
and located outside the range $7 < R < 10$ kpc, in order to minimize
contamination from the halo and outer-disk components.

Finally, we perform a simple check on the likely remaining halo contamination in
our sample following the prescription of Bensby et al. (2003). For calculation
of the approximate disk and halo star fractions (assuming our data is
representative of the local solar neighborhood densities), we adopt the local
stellar densities, velocity dispersions in $U$, $V$, and $W$, and the asymmetric
drifts listed in their Table 1, assuming the space velocities of the thin-disk,
thick-disk, and halo stars are distributed as Gaussians. Based on these
probability distributions, we reject stars that have greater likelihood of
belonging to the halo than to the disk system. This check removes only about 59
additional stars from the sample, showing that the above selection criteria for
thin- and thick-disk stars are quite reasonable. We also experimented with the
application of slightly different scale heights for describing the variation of
halo stellar densities with \z, but the above result appears quite robust. Note,
however, that these various cuts do not necessarily eliminate contamination by
members of the so-called metal-weak thick-disk (MWTD) population, which Carollo
et al. (2010) have shown exhibits metallicities in the range $-$1.7 $<$ [Fe/H]
$< -0.7$, and a prograde rotation of \vphi\ $\sim$ $+100$ to $+150$ \kms.  We
comment on any evidence for MWTD contamination below.

Summarizing the criteria used for our sample selection, surviving program stars
satisfy $d < 3$ kpc, \logg\ $\geq 4.2$, S/N $\geq 30$ \AA$^{-1}$, \vphi\ $> +50$ \kms,
[Fe/H] $> -1.2$, $7 < R < 10$ kpc, and possess greater probability of belonging
to the disk system than to the halo. The surviving sample from the above cuts
numbers $\sim$17,300 stars. Figure \ref{fig:sample} shows the distributions of
[Fe/H], \vphi, \z, and $R$ for the final dwarf sample (solid lines), before and
after further division based on the derived \afe\ ratios into the thin- and
thick-disk populations, as described below. Note that in the bottom left panel
of Figure \ref{fig:sample} has only a small number of stars with $|Z| < 0.2$
kpc, owing to the bright limit of SDSS imaging ($g > 14.0$). Thus, our analysis
in the following sections may be only valid for the thin-disk population with
$|Z| > 0.2$ kpc, rather than the young(er) thin disk closer to the plane.

\begin{figure}
\centering
\plotone{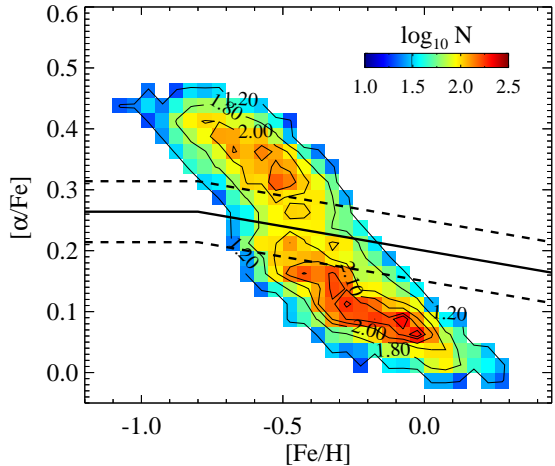}
%\plotone{contour.eps}
\caption{Distribution of logarithmic number densities, in the \afe\ vs. \feh\ plane,
over-plotted with equidensity contours. Each bin is 0.025 dex in \afe\ by 0.05
dex in [Fe/H], and is occupied by a minimum of 20 stars. The median occupancy is
70 stars. The solid line is the fiducial for division into likely thin- and
thick-disk populations; the dashed lines located $\pm$0.05 dex in \afe\ on
either side of the solid line indicate the adopted dividing points for the
high-\afe\ (upper-dashed) and low-\afe\ (lower-dashed) stars in our sample.}
\label{fig:contour}
\medskip
\end{figure}
%\newpage

\section{Division of the Sample on \afe\ into Thin- and Thick-Disk Populations}
\subsection{Dividing Scheme and its Efficiency}

As mentioned previously, since a stellar population's kinematics and spatial
distributions can be modified over time (especially in the disk system), while a
(dwarf) star's atmospheric chemical abundance is essentially invariant (except
in unusual circumstances, such as binary mass transfer from an evolved
companion), we make use of the estimated \afe\ ratio as a reference to separate
the thin- and thick-disk populations. This choice is also strongly motivated by
the apparent bi-modal distribution of stars in the 
\afe\ and \feh\ plane seen in Figure \ref{fig:contour}.

For the purpose of the present analysis, our dwarf sample is split into likely
thin-disk (with low \afe) and thick-disk (with high \afe) populations, based on the
following scheme:
\\

\noindent I) For stars with [Fe/H] $\geq -0.8$
\begin{itemize}
\item thin disk, if \afe\ $< -0.08\cdot$[Fe/H] $+~0.15$
\item thick disk, if \afe\ $> -0.08\cdot$[Fe/H] $+~0.25$
\end{itemize}

\noindent II) For stars with [Fe/H] $< -0.8$
\begin{itemize}
\item thin disk, if \afe\ $< +0.214$
\item thick disk, if \afe\ $> +0.314$
\end{itemize}

This division into the thin- and thick-disk populations is devised based on
examination of the distribution of number densities in the \afe\ vs. \feh\ plane,
shown in Figure \ref{fig:contour}. Note how well the populations appear to
separate above and below the solid line in this figure, which is our adopted
fiducial.

\begin{figure}
\centering
\plotone{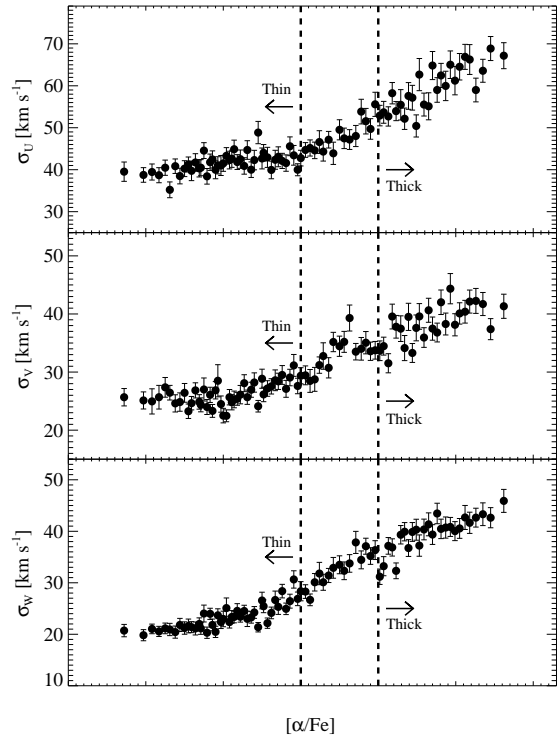}
%\plotone{afe_vel_disp.eps}
\caption{$U,~V,~W$ velocity dispersions, as a function of \afe, are shown
from top to bottom, respectively.
Well-defined trends of the velocity dispersions can be seen, providing kinematic
confirmation that the proposed separation of the thin-disk
and thick-disk populations on the basis of \afe\ works very well. Each dot
contains 200 stars, and the error bar on each point is calculated by
resampling these 200 stars with replacement 1000 times. The vertical dashed lines
provide references at \afe\ $= +0.2$ and $+0.3$, which roughly correspond to
the divisions shown in Figure \ref{fig:contour}.}
\label{fig:veldisp}
\end{figure}
%\newpage

The dashed lines located $\pm$0.05 dex in \afe\ above and below the fiducial
solid line in Figure \ref{fig:contour} indicate the dividing points for the
high-\afe\ (thick-disk) and low-\afe\ (thin-disk) stars. Note that this leaves a gap
of 0.1 dex in \afe\ between the thin- and thick-disk dividing lines. This choice
serves to reduce the number of misclassified stars that may arise from
observational errors in their measured \afe. The dashed red line in Figure
\ref{fig:sample} shows the thin-disk subsample, whereas the dotted-dash blue line is for
the thick-disk subsample, classified by the dividing schemes described above.
From this figure, one can roughly read off the ranges and peak values of the
estimated and derived parameters for each subsample.

\begin{figure*}
\centering
\plottwo{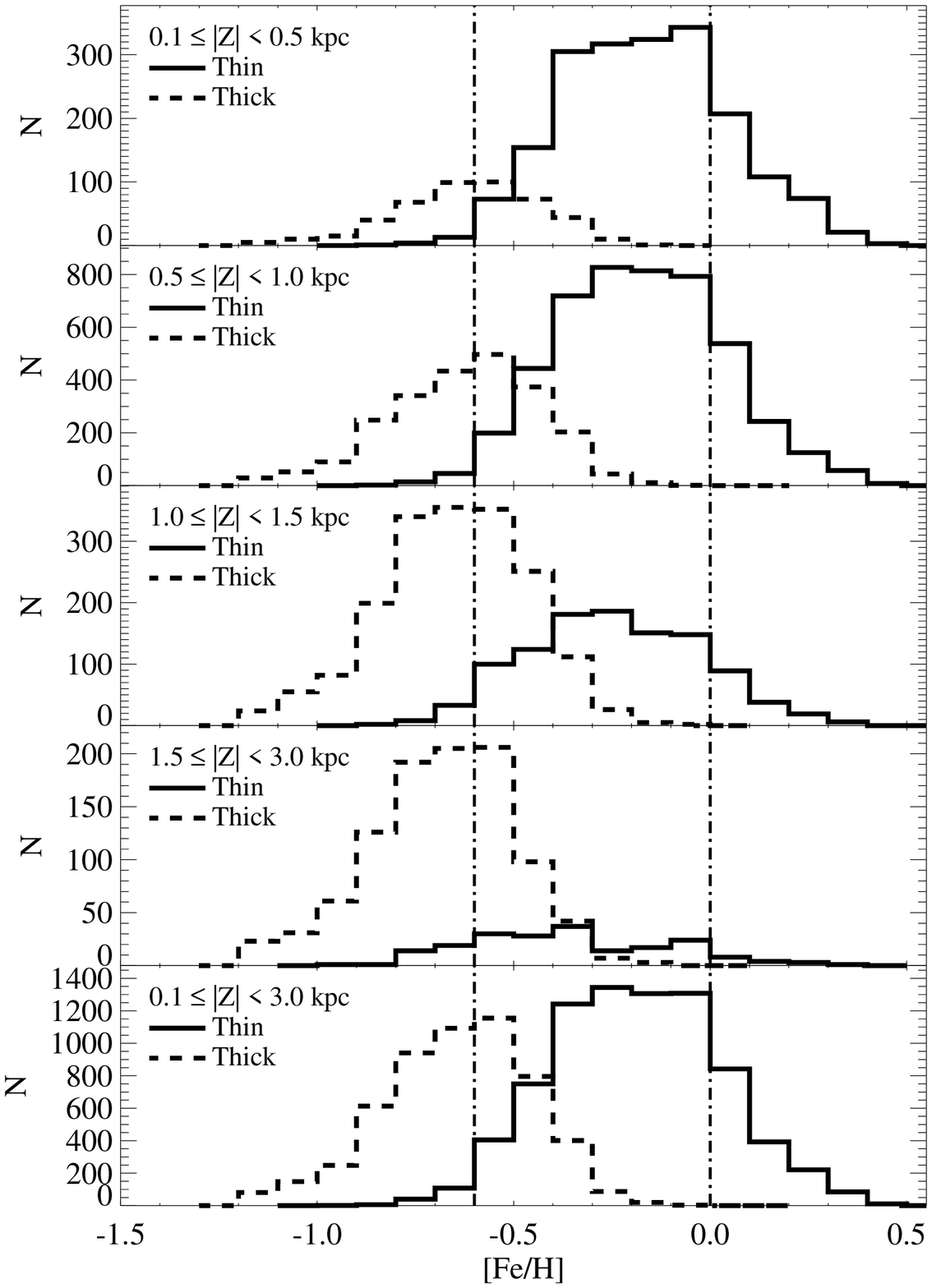}{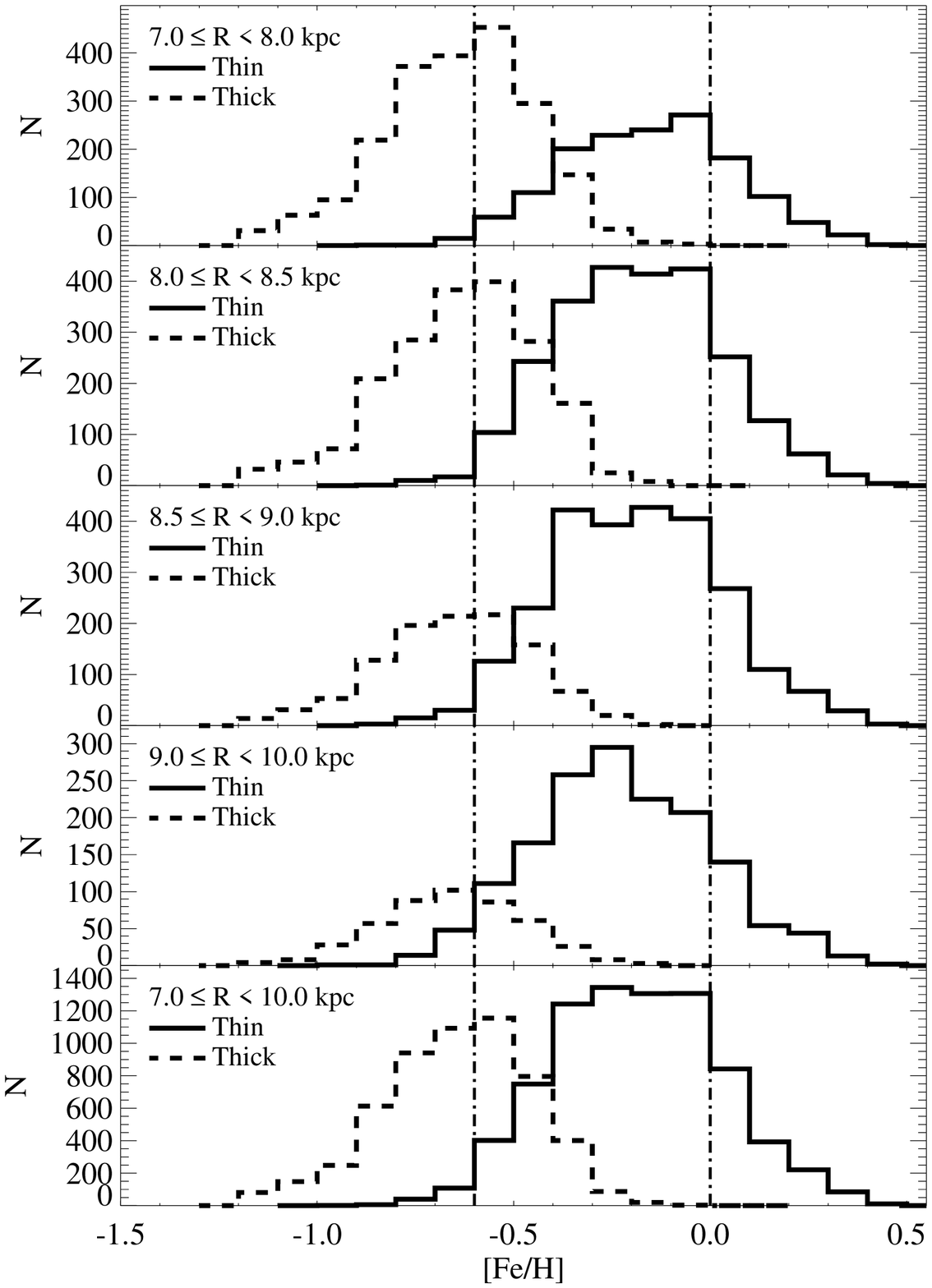}
%\plotone{bias_check.eps}
%\includegraphics[scale=0.8]{bias_check_all.eps}
\caption{Distributions of metallicities in different regions of distance
from the Galactic plane (left panel) and the Galactic center (right panel) 
for the thin-disk (solid line) and thick-disk (dashed
line) subsamples. For the thin-disk subsample, as the \z\ distance increases, the
relative numbers of metal-poor stars slightly increases (in particular for the
region \z\ $> $ 1.5 kpc). For the thick-disk subsample, the metallicity
distribution does not change with height above or below the Galactic midplane.
The metallicity distribution mostly stays the same throughout the different cuts
of Galactocentric distance for both the thin- and thick-disk subsamples. The
bottom panels show the full samples of thin- and thick-disk stars, over the
range $0.1 \le $ \z\ $ < 3.0 $ kpc. Note that the y-axis scale is very different
for each panel. The vertical dash-dotted lines are reference lines at [Fe/H] =
0.0 and [Fe/H] = $-$0.6.}
\label{fig:bias}
\end{figure*}
%\newpage

To check on the efficacy of the chemical separation of the disk
populations through use of the \afe\ ratio, we have investigated the variation of
the $U,~V,~W$ velocity dispersions of our sample with \afe. It is well known that the
dispersion of each velocity component increases with distance from the Galactic
plane (as well as on age, on average). In any event, the thick-disk population
exhibits substantially higher dispersions than the thin-disk counterpart. Figure
\ref{fig:veldisp} shows the derived velocity dispersions of our sample as a
function of \afe. It is readily apparent that, up to around \afe\ = $+0.2$, the
dispersion of each velocity component increases moderately. Above \afe\ = $+0.2$
the gradients of the velocity dispersions with \afe\ become somewhat steeper.
Above \afe\ = $+0.3$, the magnitude of each velocity dispersion is larger by
about 10 \kms\ than for \afe\ $< +0.2$. As our thin-disk stars mostly have \afe\
$< +0.2$ and thick-disk stars possess \afe\ $> +0.3$, Figure
\ref{fig:veldisp} kinematically confirms that the division by \afe\ into the
thin- and thick-disk populations is quite robust.

\subsection{Impact of Potential Metallicity Bias}

One may be concerned about biases in our initial sample which might arise from
our $(g-r)_{0}$ target selection due to small, but non-zero, metallicity
sensitivity of the stellar colors. Haywood (2001), for example, has previously
suggested that targeting stars on the basis of their spectral types (e.g.,
G-dwarfs) preferentially selected more metal-poor than metal-rich stars,
resulting in a metallicity distribution shifted lower by about 0.1--0.3 dex. 

If our selected sample strongly favors metal-poor over metal-rich stars, this
bias might produce misleading correlations between the parameters we are seeking
to understand. For example, at least for the thick-disk population, previous
studies have indicated that the observed stellar orbital rotational velocity
decreases with declining metallicity. Thus, if biases have increased the
relative numbers of metal-poor stars in the thick-disk subsample, the overall
distribution of \vphi\ will be shifted to lower rotational velocity. However, it
should be kept in mind that, because our sample does not suffer from kinematic
bias, any correlations that we are seeking between kinematics and chemical
abundances will not be affected by any potential metallicity bias, as long as
the correlations are derived from ranges of $R$ and \z\ that are sufficiently
small that the correlations remain roughly constant over the regions considered.
Furthermore, any kinematic trends with $R$ and \z\ will not be affected by
metallicity bias, as long as the metallicity distributions of stars in
different ranges of the spatial cuts do not vary significantly. 

Hence, instead of correcting for possible selection bias in our sample (which
itself is a complex and tricky business), we instead seek to demonstrate that
any presumed metallicity bias does not greatly impact the kinematic correlations
with spatial parameters that we derive by examination of the metallicity
distribution functions (MDFs) for both the thin- and thick-disk subsamples in
various regions of $R$ and $|Z|$. If there is any sudden change in the MDFs of
each subsample between neighboring regions, that will be a sign that the
subsample may not be suitable for deriving meaningful correlations between
kinematics and distance.

For the thick disk, there is some existing evidence for the lack of a
metallicity gradient with distance above the Galactic plane (e.g., Gilmore et
al. 1995), or at most for only a small one, on the order of 0.1--0.2 dex
kpc$^{-1}$ (Ivezi\'c et al. 2008). So, the shape of the observed MDFs at different 
heights should remain roughly constant. The left panel of Figure \ref{fig:bias} 
displays the observed MDFs for both the thin- and thick-disk subsamples in 
different bins of \z\ distance. From inspection, the relative numbers of 
metal-rich stars in the thick-disk subsample
do not grossly change with different cuts on height above the plane, so 
false kinematic trends with $|Z|$ are not expected to arise. Quantitatively, the
fraction of the stars with [Fe/H] $< -0.6$ for the thick-disk subsample is 0.51,
0.51, 0.58, and 0.64 from the first to the fourth panel (a resulting metallicity
gradient with \z\ of $-0.029 \pm 0.019$ dex kpc$^{-1}$), consistent with the
expectation from previous work. These results are also in line with the findings
of Haywood (2001), who noted that the metallicity bias arising from the color
selection was significant for stars with [Fe/H] $> -0.2$, but not for relatively
metal-poor thick-disk stars.

However, this seems not to be the case for the thin-disk subsample shown in the
left panel of Figure \ref{fig:bias}. At heights above \z\ = 1.5 kpc (already
many thin-disk scale heights above the plane), we notice that more metal-rich
stars have dropped out of the distribution, compared to the upper three panels.
This is quantitatively confirmed by examination of the fraction of stars with
[Fe/H] $< -0.2$ (0.45, 0.46, 0.58, and 0.71 from the first to the fourth panel).
This may be a natural consequence of selecting the sample without consideration
of the \z\ distance, since at greater heights thick-disk stars are expected to
dominate. In other words, some of the stars in the metal-poor tail of the
thin-disk subsample may in reality belong to the thick disk, but they have been
misclassified as thin-disk stars due to errors in the estimated \afe. Indeed,
considering the distribution of \afe\ for the stars with \feh\ $< -0.3$ and \z\
$> 1.5$ in the thin-disk subsample, many of the stars have \afe\ $> +0.15$.
Thus, moderately \alp-enhanced stars at this height may mostly belong to the
thick disk rather than to the thin disk, with a much lower probability of old
thin-disk membership.

Simple experiments support the above argument. According to Lee et al. (2011),
the error in \afe\ at S/N = 30 \AA$^{-1}$ is about 0.08 dex. If we assume this
is a reasonable estimate of a 1$\sigma$ random error in \afe\ for all stars with
\z\ $> 1.5$ kpc, and simultaneously perturb the measured \afe\ and \feh\ values
according to a normal distribution with $\sigma = 0.08$ dex and the measured
uncertainty in [Fe/H], the total number of stars classified as members of the
thin-disk component falls to about 85, after averaging the results of 1000
different realizations. This is substantially smaller than the 201 stars that
are claimed to be present. Thus, it is valid (within statistical fluctuations)
to say that the low-\afe\ metal-poor thin-disk stars in this \z\ distance region
are likely spurious, and are found at roughly the expected level of
contamination. Moreover, as the total number of thin-disk stars in this most
distant region is rather small, we expect the impact of such stars on our
analysis to be minimal.

Concerning the variation of the MDFs for cuts in the $R$ distance, many previous
studies (e.g., Nordstr\"om et al. 2004; Holmberg et al. 2007; Andrievsky et al.
2004; Lemasle et al. 2007; Sestito et al. 2008; Magrini et al. 2009) that used
open clusters and/or field stars to derive radial metallicity gradients report
rather small gradients, on the order of $-0.05$ to $-0.1$ \dexkpc. Thus, we
expect the MDFs in our sample to appear similar in different bins of radial
distance from the Galactic center. The right panel of Figure \ref{fig:bias}
shows the MDFs of the thin- and thick-disk subsamples for several slices in $R$.
From inspection, the overall shapes of the MDFs remain the same through the
fourth panel, although there is a small drop in the numbers of thin-disk stars
around [Fe/H] $\sim-0.1$ in the distance range $9.0 < R < 10.0$ kpc.
Calculating the fraction of the stars with [Fe/H] $< -0.6$ for the thick-disk
subsample, we obtain 0.56, 0.54, 0.58, and 0.61 from the first to the fourth
panel, a resulting radial metallicity gradient of $-0.008 \pm 0.004$ \dexkpc.
For the thin-disk subsample, the fraction of the stars with [Fe/H] $< -0.2$ is 
0.42, 0.47, 0.48, and 0.57 from the first to the fourth panel. The derived
radial metallicity gradient is $-0.043 \pm 0.004$ \dexkpc; these slopes do not
differ significantly from the previous studies. Thus, the results of these two
exercises suggest that it is highly unlikely that any potential metallicity bias
in our sample can greatly affect our derived kinematic correlations with $R$ and
$|Z|$.

Potential biases also might depend on the age distribution of our sample, which
is not known at present, and in any case is difficult to quantify. As the narrow
color range applied to originally select the G-dwarfs for spectroscopic
follow-up in SEGUE also preferentially selects certain age ranges on the main
sequence, we might expect that this bias might contribute at some level to the
observed trends (e.g., rotational velocity versus metallicity) that we are
seeking to understand. However, as noted by Haywood (2001), age bias is expected
to be even less important than the metallicity bias that we have already shown
to have minimal effect.

\section{Results of the Observations}

In this section we use our local G-dwarf sample to examine the observed
gradients of \vphi\ with \feh, $R$, and \z, as well as trends of $e$ with \feh,
$R$, and \z\ for the thin-disk and thick-disk populations as identified above.

\subsection{Correlations between Rotational Velocity and Metallicity}

\begin{figure}
\centering
\plotone{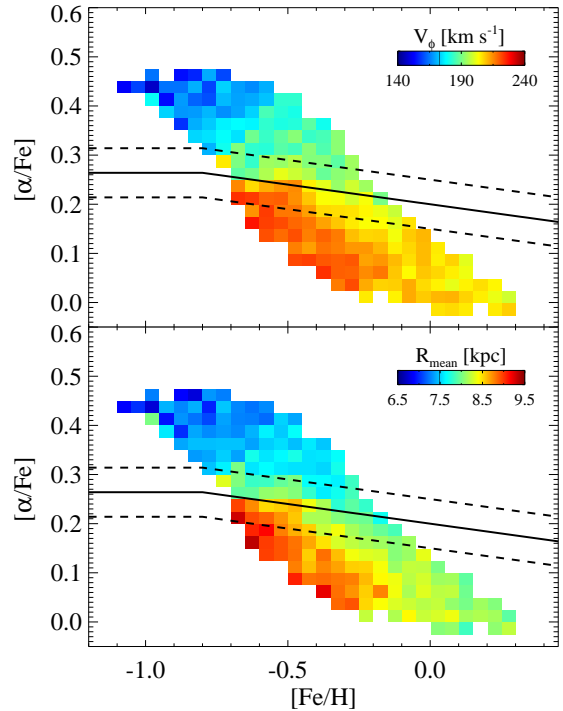}
%\plotone{color_map_afe.eps}
\caption{Distribution of rotational velocities (\vphi, top panel) and the average orbital
radii ($R_{\rm mean}$, bottom panel) for our G-dwarf sample in the \afe\ vs.
\feh\ plane. As in Figure \ref{fig:contour}, the dividing lines for the thin- and
thick-disk subsamples are shown. Each bin has a size of 0.025 dex in
\afe\ by 0.05 dex in [Fe/H], and is occupied by a minimum of 20 stars. The median
occupancy is 70 stars. Each bin represents a 3$\sigma$-clipped mean
of \vphi\ so that outliers in each bin do not significantly affect the average behavior.}
\label{fig:fehalp}
\end{figure}

The top panel of Figure \ref{fig:fehalp} shows a color-coded distribution of
\vphi\ in the \afe\ vs. \feh\ plane for our G-dwarf sample. Detailed examination
of this panel (as well as Figure \ref{fig:contour}) reveals a metal-poor tail
for the low-\afe\ stars ($< +0.2$), which we associate with the thin disk,
extending down to [Fe/H] = $-$0.7. This already implies that the thin disk may
not be well-described by a single metal-rich population with a peak around
[Fe/H] = $-$0.2. We also notice from this panel that a higher rotational
velocity is observed in the region of the metal-poor thin disk (\afe\ $< +0.2$
and [Fe/H] $< -$0.3), suggesting a negative trend of \vphi\ with \feh. In
contrast, the high-\afe\ stars ($> +0.3$, which we associate with the thick
disk) apparently exhibit a strong positive trend of \vphi\ with [Fe/H]. We
investigate these trends quantitatively below.

%\newpage

The bottom panel of Figure \ref{fig:fehalp} displays the distribution of mean
orbital radii ($R_{\rm mean}$) in the \afe\ vs. \feh\ plane. It is clear
that the stars we associate with the thick-disk population exhibit smaller mean
orbital radii than those with associated with the thin disk. In addition, the 
metal-poor thin-disk stars possess larger mean orbital radii
than the dominant metal-rich thin-disk stars.

Looking at the results from other observational work, a recent study by Navarro
et al. (2010) obtained a slightly different result for their thin-disk subsample.
These authors found little or no correlation between \vphi\ and [Fe/H] for their
thin-disk stars (defined by [Fe/H] $> -$0.7 and \afe\ $< +0.2$), although the
subset of their thin-disk subsample with available Eu abundances (so that potential
thick-disk or halo stars could be rejected) exhibits a very similar pattern to
that we identify here. One should also keep in mind the possibility of effects
from selection biases in their sample, as it was based on an assembly of stars
that included kinematically selected targets.

Haywood (2008) separated thin-disk stars with [Mg/Fe] $< +0.2$ from thick-disk
stars with [Mg/Fe] $> +0.2$ in the spectroscopic sample of Soubiran $\&$ Girard
(2005), and found an increasing trend of the mean orbital radii with decreasing
metallicity for the thin-disk population, along with a decreasing trend of mean
radii with decreasing metallicity for the thick-disk population (see his Figure
3). He claimed that this tendency resulted from stars that migrated from the
inner and outer disk. These behaviors are qualitatively in very good agreement
with our findings in the bottom panel of Figure \ref{fig:fehalp}.

\begin{figure}
\centering
\plotone{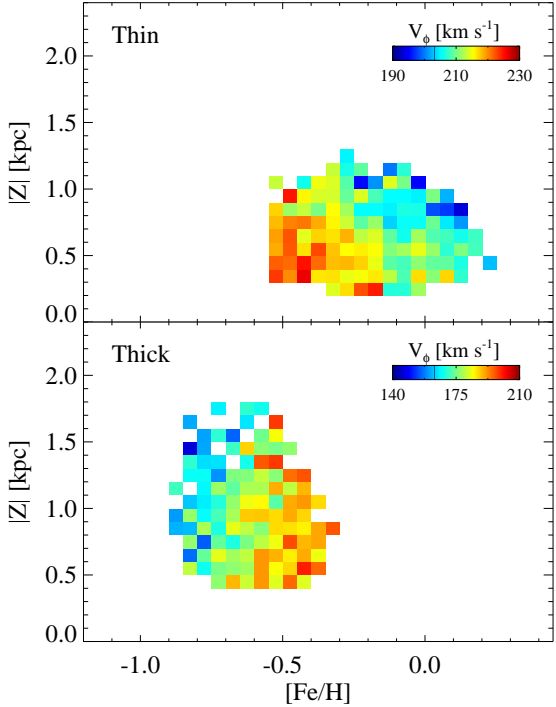}
%\plotone{color_map_zd.eps}
\caption{Distribution of rotational velocities for the low-$\alpha$, thin-disk
subsample (top panel) and the high-$\alpha$, thick-disk subsample (bottom panel) in
the \z\ vs. \feh\ plane. As in Figure \ref{fig:fehalp}, each bin has a size
of 0.025 dex in \afe\ by 0.05 dex in [Fe/H], and is
occupied by a minimum of 20 stars. The median occupancy is 47 stars.
Each bin represents a 3$\sigma$-clipped
mean of \vphi\ so that outliers in each bin do not significantly affect the
average behavior.}
\label{fig:fehzd}
\end{figure}
%\newpage

Rocha-Pinto et al. (2006) also reported a similar behavior between the mean
orbital radii and the chemical abundances in their volume-complete sample of 325
late-type dwarfs. Their primary results were that, as the difference in the
distance between the mean orbital radius and the solar radius increases, the
abundances of Fe, Na, Si, Ca, Ni, and Ba all decrease. This relationship between
the chemical abundances and the mean orbital radii could be accounted for by
radial displacements of the stars involved.

It is quite remarkable that all of the observed behaviors of the mean orbital
radii from our G-dwarf sample agree so well with several previous
observational studies (based on much smaller samples).

\begin{figure}
\centering
\plotone{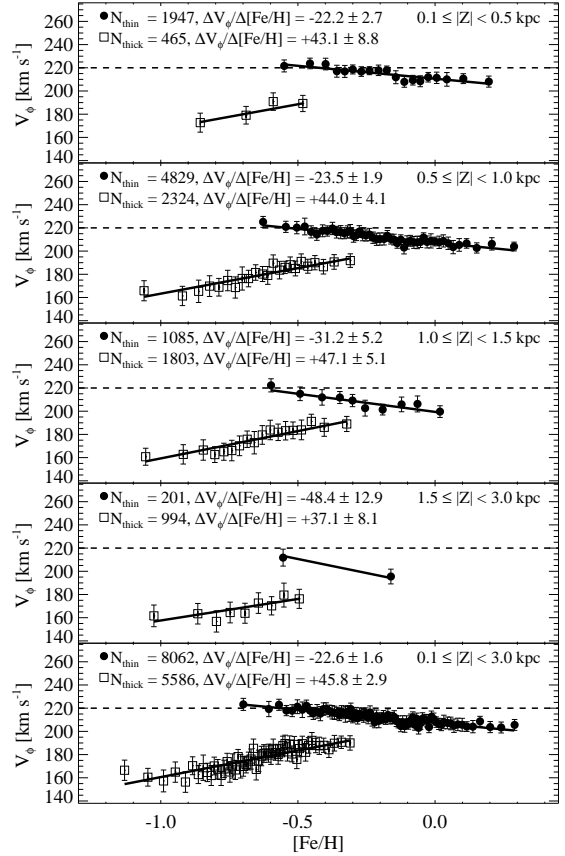}
%\plotone{feh_vphi_zd.eps}
%\includegraphics[scale=0.5]{feh_vphi_zd_mean.eps}
\caption{Rotational velocity gradients with metallicity for different slices in
distance from the Galactic plane, for stars assigned to the thin-disk (black dots)
and thick-disk (open squares) populations. Each dot represents a
3$\sigma$-clipped average of 100 stars. Note that the 1$\sigma$ error bars associated with each point
are very small (on the order of 3--5 \kms), so for visualization purposes
2$\sigma$ error bars are plotted instead. The error bars are calculated by
resampling the 100 stars in each (with replacement) 1000 times. The bottom panel
shows the results for the full samples of stars considered. Note that although
binned data are shown for clarity, estimates of the slopes and their errors are
obtained for the full unbinned data (see text).}
\label{fig:fehvphizd}
\end{figure}
%\newpage

Figure \ref{fig:fehzd} indicates that there exists a clear gradient of
\vphi\ with \feh\, at any given \z\ distance, for both the thin-disk subsample (top panel) and
the thick-disk subsample (bottom panel). Figure \ref{fig:fehvphizd} displays the
observed gradients at different heights above the plane for both subsamples.
Similar slopes of \vphi\ for both the low- and high-\afe\ stars are obtained for
the various slices in \z\ distance, although the slope of the thick-disk
subsample becomes shallower at larger distance (fourth panel), and slightly
steeper for the thin-disk subsample (which only includes 201 stars). The slopes
are obtained by performing least-squares fits to the {\it unbinned} samples of
each population; the uncertainties in the slopes are calculated by resampling
each population with replacement 1000 times. in that process, the [Fe/H] and 
\vphi\ values are perturbed simultaneously by draws from a normal distribution
using the 1$\sigma$ errors in each quantity. The data shown in the kinematic
correlation plots are binned only for clarity.

If significant contamination of our thick-disk subsample by unrecognized MWTD
stars were present, we might expect the slope of the correlation of \vphi\ with
[Fe/H] to {\it increase} with distance above the plane, due to the greater
velocity lag, larger scale height, and lower metallities of the MWTD component
compared with that of the canonical thick disk (Carollo et al. 2010). That is,
at larger distances from the plane and at lower [Fe/H], the mean \vphi\ would be
expected to be {\it lower} than it would be for a pristine thick-disk sample. We
see no evidence for steepening of the gradient in Figure \ref{fig:fehvphizd}. 
Note that this is not to be taken as a contradiction with the Carollo et al.
(2010) results, as those considered a different sample of stars, most of which
were substantially lower metallicity than considered in the present study, and
explored larger heights above the plane.

A gradient of about $-20$ to $-30$ \kmsdex, on average, is shown to exist for the thin-disk
subsample, and a strong gradient of $+40$ to $+50$ \kmsdex\ for the
thick-disk subsample. This result for the thick-disk population agrees with the
claim of Spagna et al. (2010), who derived a similar slope using F-, G-, and
K-type dwarfs from SDSS DR7. However, this finding clearly contradicts the
results of Ivezi\'c et al (2008), who found little correlation between \vphi\
and \feh. We discuss a possible resolution to this discrepancy in the Appendix.

In order to check how uncertainties in the parameters, \vphi, \feh, \afe, and
\z\ affect our derived gradients of \vphi\ over \z\ for both the
thin- and thick-disk subsamples, we have performed a simple Monte Carlo
experiment. Assuming a normal distribution with width set by the 1$\sigma$
estimated error for each parameter, random changes in each parameter were
applied to 1000 realizations of each subsample (over the full range in \z). We
obtained average gradients of $-19.5 \pm 1.0$ \kmsdex\ for the thin-disk
subsample and $+43.4 \pm 1.8$ \kmsdex\ for the thick-disk subsample, in good
agreement with those shown in the bottom panel of Figure
\ref{fig:fehvphizd}. We also have not found any notable correlations among 
the uncertainties in the parameters, which could possibly affect the derived
gradients. Thus, we believe that our derived gradients of
\vphi\ over \feh\ are not grossly affected by errors in the derived parameters.

\subsection{Rotational Velocity Gradients with Distance from the Galactic Center
and Galactic Plane}

\begin{figure}
\centering
\plotone{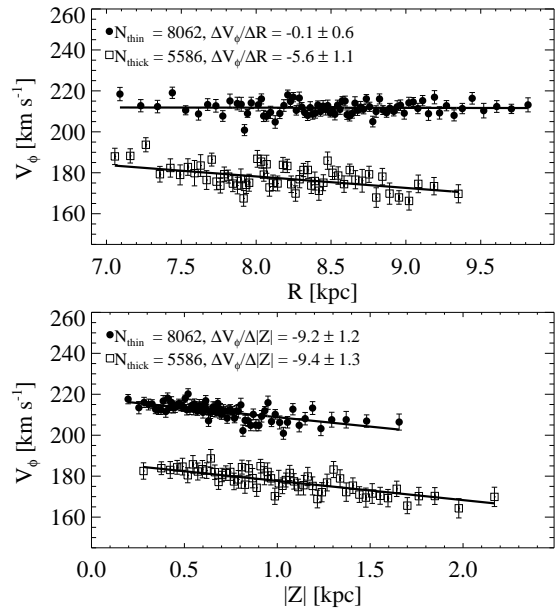}
%\plotone{r_z_vphi.eps}
\caption{Rotational velocity gradients with Galactocentric radius (top panel)
and with height above the Galactic plane (bottom panel), for the thin-disk
(black dots) and thick-disk (open squares) subsamples. As in Figure
\ref{fig:fehvphizd}, each dot represents a 3$\sigma$-clipped average of 100
stars, with error bars calculated by resampling 100 stars (with replacement)
1000 times. Note that although binned data are shown for clarity, estimates of
the slopes and their errors are obtained for the full unbinned data (see text).}
\label{fig:rzvphi}
\end{figure}

Figure \ref{fig:rzvphi} shows the overall trends of rotational velocity with
distance from the Galactic center (top panel) and with vertical distance from
the plane (bottom panel) for the thin-disk (black dots) and thick-disk (open
squares) populations. Inspection of the top panel of this figure indicates only
a negligible rotational velocity gradient for the thin-disk subsample (only
$-0.1 \pm 0.6$ \kmskpc), consistent with a flat rotation curve in the solar
neighborhood. The asymmetric drift is about 10 \kms\ at the solar radius, as
found by previous work (e.g., Soubiran et al. 2003). A small gradient of $-5.6 \pm 1.1$
\kmskpc\ is found for the thick-disk subsample, which lags the $V_{\rm LSR}$ by
$\sim$40 \kms, not far from the lag of 51 \kms\ obtained by Soubiran et al.
(2003). Note that even if we include in the analysis the stars with $0 < $
\vphi\ $<$ 50 \kms\ that were eliminated in our original selection, we obtain
very similar asymmetric drifts and gradients as for the case of a sample 
of stars that does not include them. The slopes and their uncertainties 
are obtained by the same way as for the gradient of \vphi\ with [Fe/H].

%\newpage

The bottom panel of Figure \ref{fig:rzvphi} shows that the gradients of \vphi\
with \z\ distance are very similar (about $-9.0$ \kmskpc) for both
the thin- and thick-disk subsamples. The difference in \vphi\ (the velocity lag)
for the high-\afe\ stars relative to the low-\afe\ stars is almost constant,
$\sim$30 \kms\ at any given \z\ distance. This again suggests that contamination
from MWTD stars is not a major issue for our thick-disk subsample.

Comparing with other recent studies, the vertical gradient of \vphi\ with \z\
for our thick-disk subsample, $-9.4 \pm 1.3$ \kmskpc, is smaller than that obtained by
Casetti-Dinescu et al. (2011), $-$25 \kmskpc, based on $\sim$4400 red clump
metal-rich thick-disk stars covering the metallicity range $-$0.6 $<$ [Fe/H] $<
+0.5$, that of Ivezi\'c et al. (2008) from their SDSS sample ($-$29 \kmskpc),
that of Girard et al. (2006), who derived a gradient of $-$30 \kmskpc\
from a sample of about 1200 red giants located in the range $|Z|$ = 1--4 kpc,
and as obtained by Chiba \& Beers (2000; $-$30 \kmskpc) for the subset of their
non-kinematically selected stars in the metallicity range $-0.8 \le$ [Fe/H]
$\le$ $-$0.6 within 2 kpc of the Galactic plane. Even if we cut our thick-disk
subsample to include only stars with [Fe/H] $>$ $-$0.6, we obtain a slope of
$-7.6 \pm 1.7$ \kmskpc, consistent, within 3$\sigma$, with that derived from the
subsample without a metallicity restriction.

\begin{figure}
\centering
\plotone{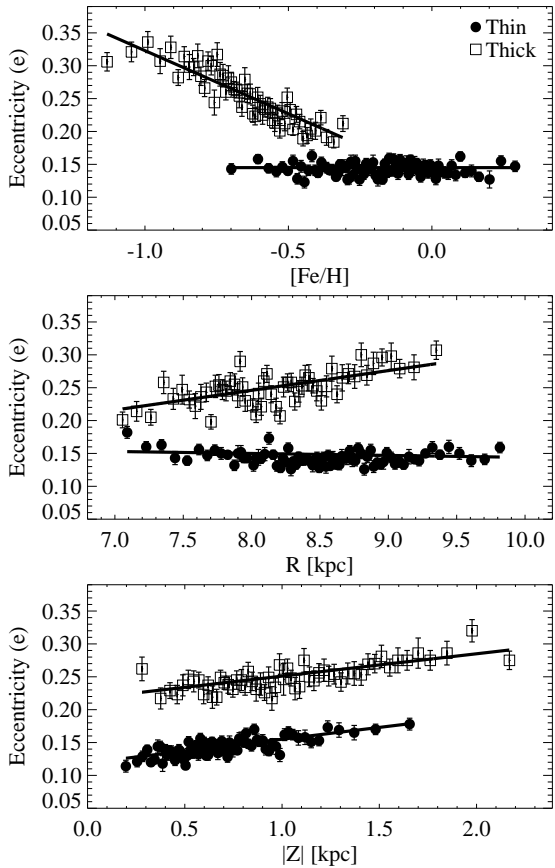}
%\plotone{ecc_param.eps}
\caption{Trends of eccentricities as a function of \feh, $R$, and \z, from the
top to bottom panels, respectively, for the thin-disk (black dots) and
thick-disk (open squares) subsamples. As in Figure \ref{fig:fehvphizd}, each dot
represents a 3$\sigma$-clipped average of 100 stars, with error bars calculated
by resampling 100 stars (with replacement) 1000 times. Note that although binned
data are shown for clarity, estimates of the slopes and their errors are
obtained for the full unbinned data (see text). The calculated gradients of $e$
and their uncertainties are listed in Table \ref{tab:obs}.}
\label{fig:eccparam}
\end{figure}

It is interesting to note that, if we consider our entire thin- and thick-disk
subsamples with $|Z| > 1.0$ kpc together, we find a vertical gradient of $-19.4 \pm 1.8$ 
\kmskpc, in better agreement with the previous studies. That is, the derived vertical
gradient of the rotational velocity becomes substantially steeper when the stars
are not divided according to their \afe\ ratios. We conclude that
accurate determination of the vertical gradient of \vphi\ with \z\ for the thick
disk {\it requires} application of a chemical separation criterion (other
than simply \feh) to isolate the various components.

Application of our simple Monte Carlo experiment with 1000 realizations of the
subsamples yielded average radial gradients of \vphi\ with $R$ of $+0.1 \pm 0.4$
\kmskpc\ and $-3.7 \pm 0.7$ \kmskpc, and vertical gradients with \z\ of $-9.2 \pm 0.9$ \kmskpc\
and $-8.2 \pm 0.9$ \kmskpc\ for the thin- and thick-disk subsamples,
respectively, without any notable covariances between the errors in the
parameters involved. These values are very close to those listed in Figure
\ref{fig:rzvphi}.

\subsection{Correlations of Stellar Orbital Eccentricities with Metallicity,
Distance from the Galactic Center, and Height Above the Galactic Plane}

Figure \ref{fig:eccparam} shows trends of orbital eccentricities ($e$) for the
G-dwarf sample, as a function of \feh, $R$, and \z, from the top to bottom
panel, respectively. The black dots denote our thin-disk subsample, while the
open squares indicate the thick-disk subsample.

One outstanding feature from inspection of the three panels is that the overall
distribution of the orbital eccentricities for the thick-disk stars is easily
separable from that for the thin-disk population. The top panel suggests that
the trend of the eccentricities for the thin-disk stars is independent of
metallicity, i.e., an almost flat trend of $e$ with [Fe/H], indicative of a very
narrow distribution of eccentricity, with a peak around $e\sim0.14$. On the
other hand, the trend of $e$ for the thick-disk subsample generally increases as
the metallicity decreases. A slope of $-0.192 \pm 0.010$ dex$^{-1}$ is obtained
from a least-squares fit to the averaged points.

The second panel also shows several interesting features. As in the top panel,
there is not much correlation between $e$ and $R$ for the thin-disk subsample,
although the behavior trends slightly higher below $R=7.5$ kpc. The
thick-disk stars generally exhibit an increasing trend of $e$ with increasing $R$.
The eccentricity distributions for the thin- and thick-disk populations merge at
$R\sim7.0$ kpc.

The bottom panel shows that the eccentricities for both low- and high-\afe\ stars
increase on average the farther away they are from the Galactic plane. In addition,
similar to Figure \ref{fig:rzvphi}, it is also noticed that the difference (about 0.1) in
the eccentricity between the low-\afe\ and high-\afe\ subsamples is constant
at any given \z.

A simple Monte Carlo experiment with 1000 realizations of the
subsamples also reveals that the derived trends of the eccentricities with
\feh, $R$, and \z\ above are not strongly affected by errors in the parameters
involved, as the computed gradients of the eccentricities are
within 3$\sigma$ from those values listed in Table \ref{tab:obs}, which
quantitatively summarizes various correlations discussed in this
section for the two subsamples.

%% Table 1
\begin{deluxetable*}{ccccccc}
\tabletypesize{\scriptsize}
\tablewidth{0in}
\renewcommand{\tabcolsep}{1pt}
\tablecaption{Summary of Observed Gradients for
the Thin- and Thick-disk Subsamples}
\tablehead{\colhead{Correlation} & \colhead{$\Delta$\vphi$/$$\Delta$\feh} & \colhead{$\Delta$\vphi$/$$\Delta$$R$} &
\colhead{$\Delta$\vphi$/$$\Delta$$|Z|$} & \colhead{$\Delta$$e/$$\Delta$\feh} &
\colhead{$\Delta$$e/$$\Delta$$R$} & \colhead{$\Delta$$e/$$\Delta$\z} \\
\colhead{} & \colhead{(\kmsdex)} & \colhead{(\kmskpc)} &
\colhead{(\kmskpc)} & \colhead{($\rm dex^{-1}$)} & \colhead{($\rm kpc^{-1}$)} & \colhead{($\rm kpc^{-1}$)}}
\startdata
Thin & $-22.6\pm1.6$ & $-0.1\pm0.6$  & $-9.2\pm1.2$ &$+0.000\pm0.005$ &$-0.003\pm0.002$ & $+0.036\pm0.003$\\
Thick & $+45.8\pm2.9$ & $-5.6\pm1.1$ & $-9.4\pm1.3$ &$-0.192\pm0.010$ &$+0.030\pm0.004$ & $+0.034\pm0.004$
\enddata
\label{tab:obs}
\end{deluxetable*}

\section{Qualitative Comparisons with Predictions of Contemporary Models of Disk Formation}

It may be unwise to rely too strongly on the present predictions of the
suggested thick-disk formation models. This follows because, even though they
are able to reproduce some aspects of the Milky Way's disk system, the predicted
properties are limited by large uncertainties with their treatment of star
formation, the dynamical interaction of presumed satellites with the disk,
unavoidable numerical effects, and the myriad set of assumptions that are required
in their construction. Thus, in this section, we compare our observational
findings only with qualitative expectations from the published radial migration,
gas-rich merger, accretion, and disk heating models. It is our expectation that,
as the models and simulations improve, these comparisons will increasingly be
able to discriminate between the relative importance of the various formation
scenarios.

\subsection{Correlations between Rotational Velocity and Metallicity}

According to the radial migration models (Sellwood \& Binney 2002; Ro\v skar et al. 2008a;
Sch\"onrich \& Binney 2009a; Minchev \& Famaey 2010), the (presumably) metal-poor stars
of the thin disk (which includes young, low-\afe\ stars) that were
born in the outer disk move inward to the solar neighborhood, while the
(presumably) metal-rich stars that formed in the inner disk migrate outward into
the solar neighborhood (as the inner region of the disk has a higher stellar and
gas density, and is rapidly chemically enriched, most of the stars should be
metal rich).

Sch\"onrich \& Binney (2009a) suggested that this radial movement
can occur by two mechanisms: ``blurring'' and ``churning''. Blurring refers to
the increase of eccentricities over time at a similar angular momentum due
to scattering, e.g., on giant molecular clouds. Churning is mostly triggered by
resonant scattering at co-rotation due to transient spiral density waves, which
transfers stars from inner (or outer) disk regions into the solar vicinity by changing
their angular momenta without alteration of their orbital circularity (hence
eccentricities). These authors suggested that churning is the dominant process by
which stars in the inner disk migrate out to the solar annulus, thus providing
greater heterogeneity in the abundance and velocity distributions among solar
neighborhood stars.

The consequence of incomplete mixing from blurring and churning is that
the metal-rich stars in the thin disk possess relatively
lower rotational velocities (\vphi), while the metal-poor stars have
higher \vphi. Thus, the expectation is that there should exist a
trend of \vphi\ with [Fe/H] among (at least) the thin-disk stars.
Sch\"onrich \& Binney (2009a,b) indeed predicted a significant downtrend
of \vphi\ with \feh\ for the low \afe\ stars, due to incomplete mixing for
younger stars. This prediction was confirmed by the later N-body models
of Loebman et al. (2010), who employed slightly different treatments
of radial mixing and star formation in their simulated disks from
Sch\"onrich \& Binney (2009a,b), but found a gradient of $-$19.7 \kmsdex\
for younger stars (identified with the thin-disk component with low \afe) in
the solar neighborhood ($7 < R < 9$ kpc and $0.5 <$ \z\ $< 1$ kpc).

Our observed gradient of \vphi\ with [Fe/H] for the thin-disk stars in
the range $0.5 < |Z| < 1.0$ kpc ($-23.5 \pm 1.9$ \kmsdex) is not far from the estimate
of $-$20 \kmsdex\ obtained by Loebman et al. (2010) for their simulated sample
of young, low-\afe\ stars in their $transition~zone$, which covers the same
interval in height above the plane. Note that the scale of their \afe\
determinations and ours are slightly different, and they also employed the
predicted oxygen abundance ratio as a proxy for \afe, rather than the averages
employed in our estimates. It appears that an overall velocity gradient of
$-20$ to $-30$ \kmsdex with metallicity for the thin-disk subsample
qualitatively agrees well with the expectations from the radial migration
models.

The extended tail of low-\afe\ metal-poor stars observed in Figures 1, 2, 4, and
5 can be also explained by the radial migration scenario. Ro\v skar et al.
(2008b) found that their simulated disk stars (when allowed to mix radially)
exhibited a MDF more like the observations by Holmberg et al. (2007) than that
obtained from an $in~situ$ sample without radial migration. These authors
concluded that radial migration was the likely cause of the broader MDF, which
is also supported by our present data.

It is noteworthy that our observed negative gradient of \vphi\ with [Fe/H] for
the thin-disk stars ($-20$ to $-30$ \kmsdex) stands in contradiction to
expectations from traditional local evolution models in the solar neighborhood
(without allowing for mixing or migration of stars), which predict a {\it
positive slope} of \vphi\ with [Fe/H]. According to these models, the stars that
were born early in the history of star formation in the thin disk are expected
to be relatively metal-poor. These old metal-poor thin-disk stars should have
experienced more perturbations, such as from variations in the Galactic
potential over time. As a result, such stars are expected to exhibit slower
rotational velocities and larger velocity dispersions than the younger, more
metal-rich thin-disk stars. This inevitably leads to the expected production of
a positive gradient of \vphi\ with [Fe/H], which we clearly do not find.

When considering the radial migration models for the thick disk the case
differs somewhat, in particular due to the much older ages of these stars.
According to these models, it is expected that most of the thick-disk stars that
exist in the solar neighborhood today were born with high velocity dispersion in
the inner portion of the Galaxy, in regions of higher local density, at a time
when the metallicity of the ISM was relatively low and the \alp-abundance ratios
were high. As they migrated outward over time, the lower gravitational restoring
force of the local disk allowed these stars to explore orbits reaching higher
above or below the plane. Relatively few thick-disk stars are thought to have
migrated inward from the outer disk region. These old stars had more time to
experience mixing of their orbits; in the case of complete mixing for these
older stars, one might expect little or no trends between rotational velocity
and metallicity.

Sch\"onrich \& Binney (2009a,b) did not make predictions of velocity trends
with metallicity, on the grounds that insufficient knowledge of
the earliest phases of disk formation exists to constrain expectations
for such a potential gradient (i.e., unknown initial conditions).
However, Loebman et al. (2010) reported from their simulation an insignificant
gradient of $+$1.4 \kmsdex\ for these older stars ($> 7$ Gyr, which generally
matched the observed properties of the thick-disk component, e.g., high \afe\
ratios). Even though the migration strength in their simulation induced
substantial mixing, the process was still incomplete. Thus, it would allow for
the conservation of significant velocity/metallicity trends. It should be
mentioned, however, that their model was not specifically intended to match the
properties of the Milky Way.

In any case, the small or absent predicted correlations between \vphi\ and \feh\
for the high-\afe\ stars from the migration models are in contrast to
our determination of a steep gradient of $+40$ to $+50$ \kmsdex\ for the
observed high-\afe\ stars we associate with the thick disk, as shown in Figure
\ref{fig:fehvphizd}. Thus, this trend of \vphi\ with \feh\ for the thick-disk
subsample can provide a useful constraint to the radial migration models
mentioned above.

In summary, the observed correlations between \vphi\ and [Fe/H] for our
low-\afe\ (thin-disk) stars can be naturally explained by the radial migration of
stars from the outer disk (more metal-poor stars) and from the inner disk (more
metal-rich stars) into the solar vicinity, as predicted by the migration models.
As explained by Sch\"onrich \& Binney (2009b), such a velocity gradient arises from
the interplay between the churning and the blurring processes. The behavior
of the high-\afe\ (thick-disk) stars is rather different (exhibiting a much
steeper gradient) than expected from the simulated high-\afe\ stars of Loebman et
al. (2010), although there remains the uncertainty of how well thick-disk
stars are represented in these models, and how well the models match the actual
history of the Milky Way. It appears that stellar radial migration
may have played an important role in the evolution of the thin disk, but, based
on the information available from the current radial migration models and
simulations, it is difficult to ascertain the relative importance of radial migration
for the formation and/or evolution of the thick disk.

\subsection{Rotational Velocity Gradients with Distance from the Galactic Center
and Galactic Plane}

The gas-rich merger model of Brook et al. (2007) predicts a
correlation between \vphi\ and $R$ for stars in the disk system. According to
their simulations (especially their Figure 5) there should exist a detectable
velocity gradient for the thin disk in the region of the solar neighborhood ($R$
= 7--10 kpc). This differs from our null gradient for the thin-disk subsample.
Their simulations also indicate a negligible gradient for their thick-disk stars
(which they refer to as ``merger stars''), which is at least qualitatively in
line with our small value of $-5.6 \pm 1.1$ \kmskpc. However, it should be kept in mind
that, as Richard et al. (2010) demonstrated in their various gas-rich merger
simulations, the initial orbital parameters of the mergers strongly affect the
final kinematics and structures of the resulting disk populations. Brook et al.
(2007) performed a simulation with a particular set of parameters to produce
their disk systems, which may not necessarily match those of the Galaxy. For
example, one rather large difference between this particular simulation and our
results is that, while we find a difference in velocity lag of about 30 \kms\
between our thin and thick-disk subsamples, the $N$-body prediction calls for a
difference of over 150 \kms. Additional simulations of this process, better
matched to the nature of the Milky Way, would clearly be useful to compare our
results with.

The dynamical heating of a pre-existing thin disk, as modeled by the simulations
from Villalobos et al. (2010) also predicts gradients of \vphi\ with respect to
both $R$ and \z. Looking at their Figure 14, the thickened-disk component
exhibits a very weak trend of \vphi\ with $R$ for low initial orbital
inclination of the merging satellite, while the correlation between the two
quantities becomes stronger as the incidence angle is increased. Concerning the
gradient of \vphi\ with \z, it is evident in their Figure 14 that the vertical
gradient of \vphi\ is much shallower at higher orbital inclination. Thus,
roughly speaking, our radial gradient of \vphi\ for the thick-disk subsample
agrees better with that expected for low orbital inclination of the merging
satellite, but our vertical gradient is better matched by mergers with high
orbital inclination. This may indicate that the compromise case of intermediate
orbital inclination ($i = 30^{\circ}$) best describes our observed results, a
possibility also considered by Villalobos et al. (2010). In any event, the
comparisons of our thick-disk subsample with this particular model prediction
imply that if the heating scenario played a major role in the formation of the
thick disk, the initial orbital inclination of the merging satellite could not
have been too small or too large. Of course, it is also possible that multiple
satellite mergers may have been involved, which complicates these simple
comparisons with a single merger.

\begin{figure*}
\centering
\plottwo{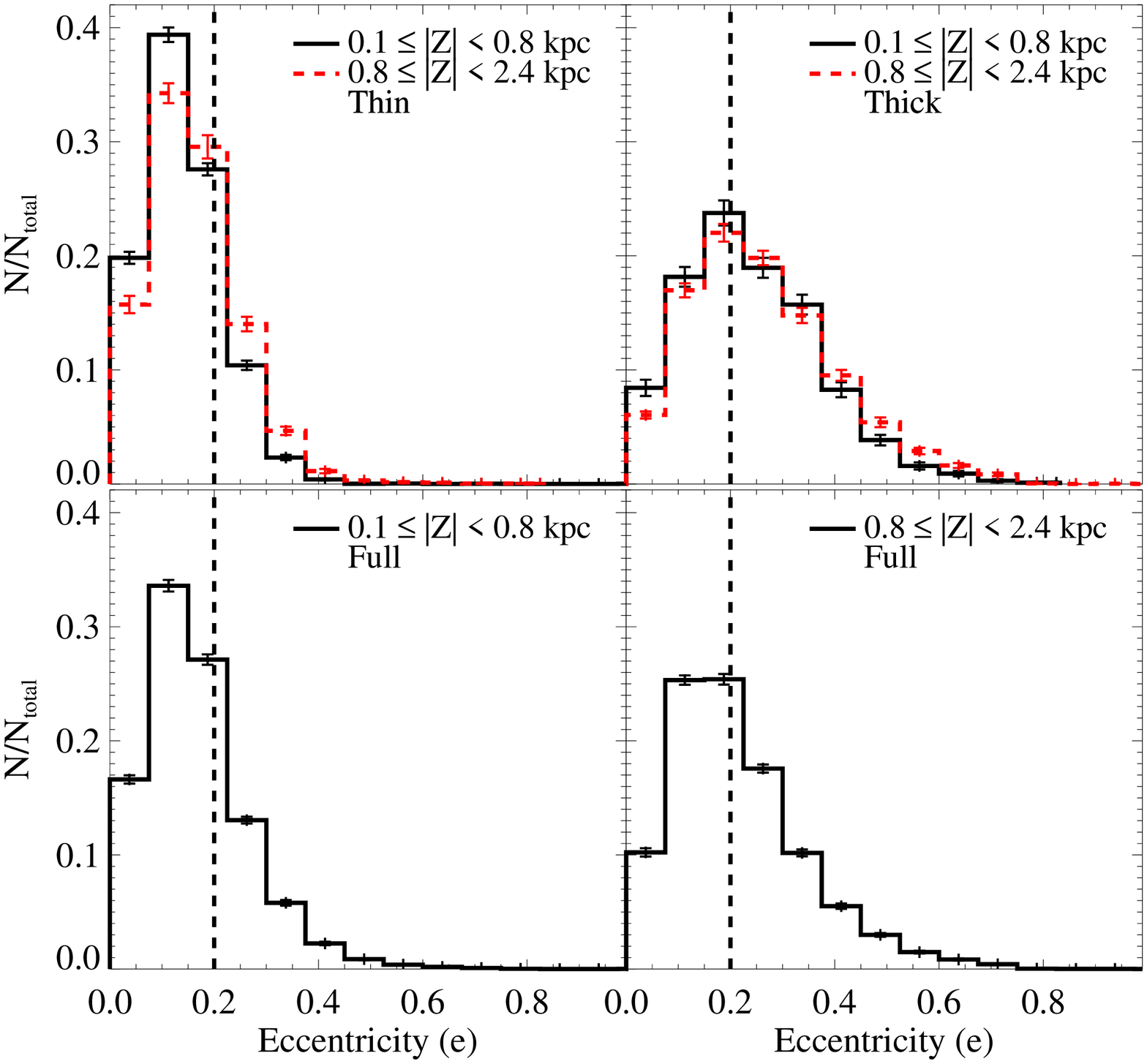}{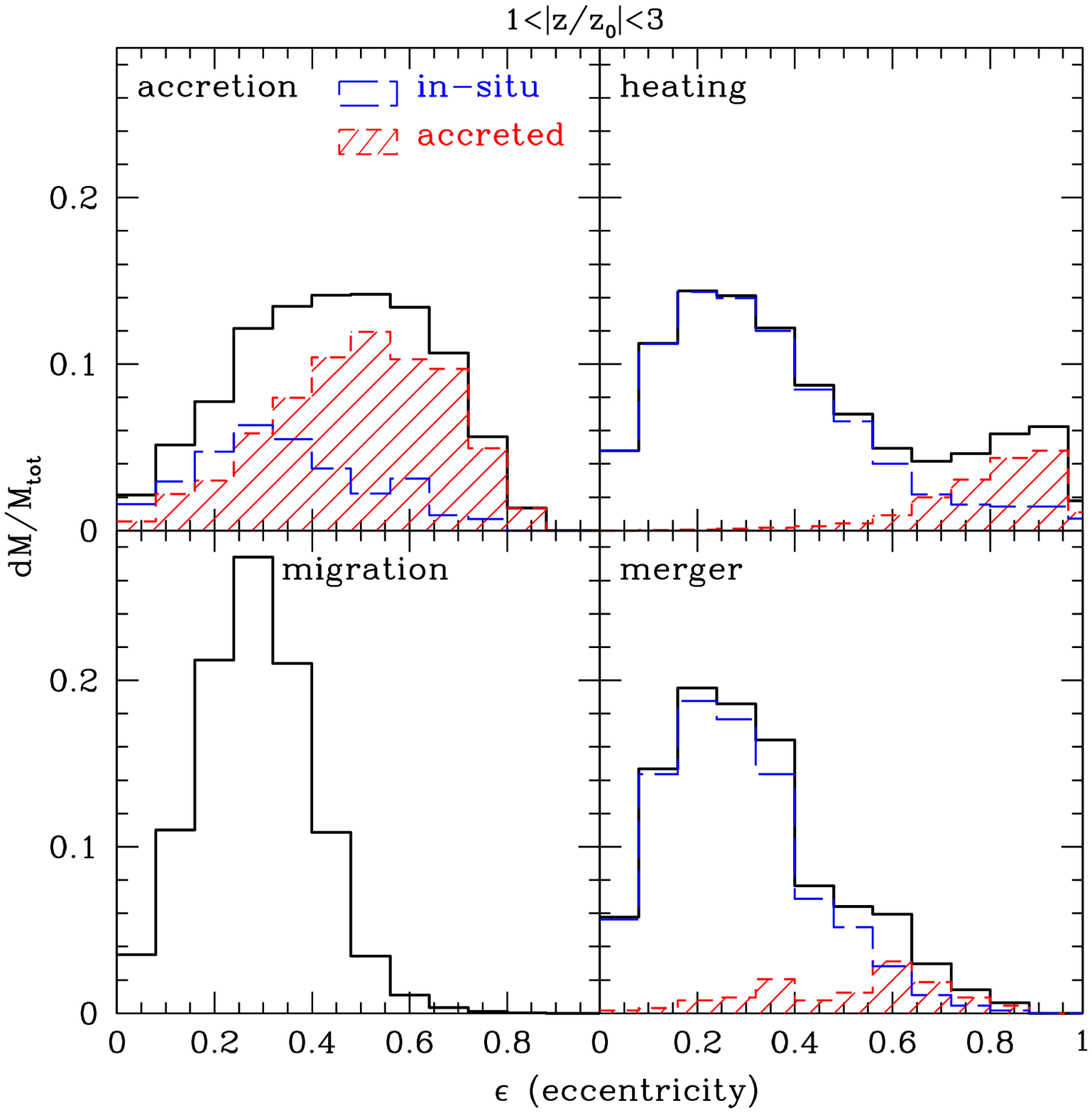}
%\plottwo{ecc_dist.eps}{e_hist_allsimus.eps}
\caption{$Left~panel:$ Normalized distributions of eccentricities for the
low-\afe\ stars (top left panel), high-\afe\ stars (top right panel),
and for the entire G-dwarf sample (two bottom panels) over
different \z\ ranges. The vertical dashed line provides
a reference at $e = 0.2$. The error bars are calculated by drawing from the sample
with replacement 1000 times. A typical error in $e$ is 0.044.
$Right~panel:$ A reproduction of Figure 3 from Sales et al. (2009). 
The formation scenarios depicted are adopted from Abadi et al. (2003) for the
accretion model, Villalobos \& Helmi (2008) for the heating model, Ro\v skar et
al. (2008a) for the radial migration model, and Brook et al. (2004, 2005) for the
gas-rich merger model. We adopt a bin size of 0.075 in the left-hand panels,
close to that used in the Sales et al. figure. Taking the scale height ($Z_{0}$)
of the thick disk as 0.8 kpc, the range $1 < |Z/Z_{0}| < 3$ in the Sales et al.
figure corresponds to the range $0.8 < |Z| < 2.4$ kpc.}
\label{fig:eccdist}
\end{figure*}
%\newpage

In summary, comparisons of our data with the gas-rich merger model from Brook et
al. (2007) suggest that, while this model may not explain the lack of a
rotational velocity gradient with Galactocentric distance for the thin disk, it
does account for that observed for the thick disk. However, the much larger
difference in the velocity lag than our finding between the thin- and thick-disk
stars remains to be resolved. The model of thin-disk heating by mergers of
Villalobos et al. (2010) qualitatively agrees with the expected kinematic
features of our thick-disk subsample, assuming that the merging satellite has an
intermediate orbital inclination.

In the previous section, which considered correlations between
\vphi\ and \feh, our results for the thin-disk population were shown to be in
qualitative accord with predictions of the radial migration models, while those
for our thick-disk population might not be. In order to be confident of the
implications of this result, one would like to compare with the predictions from
more refined radial migration models that better reproduce the observed
properties of the thick-disk population. On the other hand, the relationship
between \vphi\ with $R$ and \z\ for the high-\afe\ stars agrees better with the
predictions of the gas-rich merger and thin-disk heating models that we have
considered here. Taken as a whole, the presently available comparisons of the
various observed gradients suggest that the thick disk may have formed from
either the mergers of gas-rich systems or the heating of a pre-existing thin
disk by mergers, and has been little influenced by the secular process of
stellar migration, while radial migration may well have strongly affected the
evolution of the thin disk.

\subsection{Distribution of Stellar Orbital Eccentricities}

Sales et al. (2009) demonstrated that the orbital eccentricities of a stellar
population could also be used as a tool to probe the formation and evolution
mechanisms of the disk system. In particular, taken at face value (and
recognizing that their summary only pertains to a limited set of model
parameters and histories), their Figure 3 suggests that radial migration models
(e.g., Ro\v skar et al. 2008a) generate symmetric distributions of stellar
eccentricities with rather narrow widths, while the gas-rich merger models
(e.g., Brook et al. 2004, 2005) produce distributions that are skewed toward
higher eccentricity with larger widths. The accretion models (e.g., Abadi et al.
2003) distribute the eccentricities rather broadly over a wide range. For the
disk heating scenario (e.g., Villalobos \& Helmi 2008), there is a similarity of
the eccentricity distribution with that of the merger model for $e < 0.6$, but
there exists a secondary peak at high eccentricity ($e\sim0.8$). Generally, they
found that violent models such as disk heating and accretion generated a
distribution of stellar orbital eccentricities spanning a large range, with
secondary peaks at higher eccentricity, or at least with rather broad
distributions of high eccentricity stars. By contrast, the smooth transition
models, such as radial migration or $in~situ$ star formation from gas-rich
mergers produced distributions dominated by lower eccentricity orbits covering
relatively narrower ranges.

Several studies have compared the above expectations from these models to
observed distributions of orbital eccentricities for thick-disk stars in the
solar neighborhood. Wilson et al. (2011), for example, investigated the
eccentricity distribution of a sample of thick-disk stars from RAVE. They
concluded that their observed distribution, which peaked at low eccentricity and
exhibited a lack of high eccentricity stars, disfavored the pure accretion
model of Abadi et al. (2003), and was most consistent with the predictions of
gas-rich merger models. Dierickx et al. (2010) carried out a similar test, using
a large sample of dwarfs from SDSS DR7, and suggested that their sample favored
the gas-rich merger scenario as well. Casetti-Dinescu et al. (2011) performed an
analysis using a sample of $\sim$4400 red clump thick-disk stars from RAVE Data
Release 2 (Zwitter et al. 2008) with available proper motions from SPM4. Their
comparison of the derived orbital eccentricity distribution with model
predictions supported the gas-rich merger scenario, or possibly the minor merger
heating model (arguing that the expected secondary peak at high eccentricity
could be avoided, depending on the initial orbital configuration of the merging
satellite(s)). Indeed, a recent simulation study by Di Matteo et al. (2011)
showed that, with the adoption of a particular set of initial conditions (a 1:10
mass ratio and direct orbit of a presumed single interacting satellite), the
disk heating model could also produce the distribution of eccentricities
observed by Wilson et al. (2011) and Dierickx et al. (2010) {\it without}
creating a secondary peak at high eccentricity, confirming that the heating
model may also be a viable mechanism for thick-disk formation.

It is noteworthy that the various observational studies, based on different
samples, with different distance estimates, and adopting different models for
the Milky Way potential, all produce similar eccentricity distributions for the
thick-disk population -- a broad peak at low eccentricity and a lack of high
eccentricity stars. Considering all of the studies mentioned above, the favored
mechanisms for thick-disk formation are likely to be either (or both) the
gas-rich mergers model or the thin-disk heating by minor mergers scenario, at
least when considering only stellar orbital eccentricities as a probe. All of
these studies rejected the pure accretion model of thick-disk formation (as
advocated by Abadi et al. 2003).

Unlike the previous observational studies mentioned above, which
selected thick-disk stars mostly on the basis of spatial extent, we have selected a
subsample of likely thick-disk stars based on their measured \afe, as described
in Section 3. We now compare the eccentricity distribution of our thick-disk
subsample with expectations from each model cited in Sales et al. (2009). The
left-hand set of panels of Figure \ref{fig:eccdist} displays the normalized
distributions of eccentricity for the low-\afe\ (top left) and high-\afe\ (top
right) populations. Each distribution in the top two panels is restricted to
different slices on distance from the Galactic plane, as listed in the figure
legend. The $e$ distribution of the entire G-dwarf sample (without splits based
on \afe), divided into regions that should emphasize the thin- and thick-disk
regions, is shown in the bottom left and bottom right panels, respectively.

The eccentricity distributions of the thin-disk subsample peak at much less than
$e=0.2$, with narrow widths, and apparently include very few high
eccentricity stars ($e > 0.4$) for the two \z\ regions shown in the top left
panel. In contrast, the distributions for the thick-disk subsample shown in
the top right panel peak at $e\sim$0.2, and exhibit extended tails of
higher eccentricities up to $e\sim$0.8; there remains a
relative lack of high eccentricity stars ($e > 0.6$). Although we find that the
relative frequency of the high-$e$ stars increases a bit at larger \z\ distance
(red dashed line) for both subsamples, the distributions otherwise do not change
significantly. This again confirms that the population separation based on \afe\
appears to work quite well. The eccentricity distributions for the full sample
of G-dwarf stars exhibits some rather interesting features. Even at large \z\
distance (0.8--2.4 kpc, bottom right), where the thick-disk stars should
dominate, the eccentricity {\it does not appear similar} to that of the thick-disk
subsample separated by \afe\ (top right panel) in either range of \z\ distance;
the peak and the width do not match. This underscores once more that, for the
purpose of the selection of thick-disk stellar samples, purely spatial
separations are insufficient.

%% Table 2
\begin{deluxetable*}{ccccc}
\tabletypesize{\scriptsize}
\tablewidth{0in}
\renewcommand{\tabcolsep}{1pt}
\tablecaption{Results of Qualitative Comparison Tests with Predictions of
Published Models for Thick-disk Formation}
\tablehead{\colhead{Model} & \colhead{$\Delta$\vphi$/$$\Delta$\feh} &
  \colhead{$\Delta$\vphi$/$$\Delta$$R$} & \colhead{$\Delta$\vphi$/$$\Delta$$|Z|$} &
  \colhead{$e$ Distribution}}
\startdata
Accretion       &  N/A    & N/A        &  N/A  & Failed \\
Disk Heating    &  N/A    & Passed     &  Passed  & Failed \\
Radial Migration&  Indecisive\tablenotemark{1} & N/A    &  N/A  & Indecisive \\
Gas-rich Mergers&  N/A    & Passed     &  N/A  & Passed
\enddata
\tablecomments{N/A indicates that a model prediction is not available.
The adopted thick-disk formation models are drawn from:  Abadi et al. (2003) for the
accretion model, Villalobos \& Helmi (2008) for the heating model, Ro\v skar et
al. (2008a) for the radial migration model, and Brook et al. (2004, 2005) for the
gas-rich mergers model.}
\tablenotetext{1}{Based on the comparisons with predictions by Sch\"onrich \& Binney (2009a, 2009b) and 
Loebman et al. (2010).}
\label{tab:test}
\end{deluxetable*}

Comparing with the published model predictions in Sales et al. (2009), as shown in the
right-hand panels of Figure \ref{fig:eccdist}, the relative shortage of high
eccentricity stars and the absence of the secondary peak at high $e\sim$0.8 in
our observed distribution exclude the accretion origin and the disk heating
model for the thick disk. Although the distribution expected from the radial
migration models provides a viable description of stars in the low eccentricity
region, it fails to capture the observed high eccentricity tail of the
thick-disk stars. The skewed distribution of observed eccentricities toward
higher values is not well-represented by the radial migration predictions, which
exhibit a more Gaussian-like shape (lower left panel of the right-hand panels
of Figure \ref{fig:eccdist}). It should be noted, however, that an alternative
radial migration model by Sch\"onrich \& Binney (2009a,b) indicates the presence
of a peak eccentricity between 0.1 and 0.2, with an extended tail towards high
eccentricities, which is consistent with the shape of the observed $e$
distribution of our thick-disk sample. Hence, we must be cautious in drawing
firm conclusions on the formation mechanisms of the thick disk due to their
apparent sensitivity to details of the models and simulations. Solely based on
comparisons with the predictions in the published models from Sales et al.
(2009), it seems that our eccentricity distribution most closely resembles that
predicted from the gas-rich merger scenario.

The eccentricity distribution of our thick-disk subsample differs little from
the disk heating model of the Di Matteo et al. (2011) simulation. As identified
by this simulation (and also mentioned in the discussions of the observations of
Casetti-Dinescu et al. 2011 and Wilson et al. 2011), the secondary peak or high
eccentricity region ($e > 0.6$) in the disk heating model is mostly occupied by
the accreted stars (which retain the initial orbital characteristics of the
merging satellite).  In addition, depending on the initial conditions
(especially the orbital inclination of the interacting satellite) of the
simulation, the high eccentricity secondary peak may (or may not) be seen in the
predicted distribution of the eccentricities. In particular,
the small satellite mass (1:10 mass ratio) in the Di Matteo et al. simulation
would likely not contribute large numbers of stars to the solar neighborhood.

It appears from our observed eccentricity distribution, and that of others, that
the inclination of the merging small galaxy in the simulation of the disk
heating model in Sales et al. (2009) may be less than 30$^{\circ}$. This is
qualitatively consistent with the findings from the correlations of \vphi\ with
$R$ and \z\ in the previous section. We stress that the existence of the
secondary peak at high eccentricity and clear identification of the extended
tail of the high eccentricity with observational data can provide strong
constraints on the initial conditions on the merger or heating models.

The observed eccentricity distribution of the full G-dwarf sample at larger
distances from the Galactic plane, or for the thick-disk population with high
\afe\ ratios, rule out a broad peak at intermediate
eccentricity. This argues strongly against the importance of an accretion origin
of the thick-disk component, unless the accretion model can explain the dominant
population in the low eccentricity regime. At this stage, confident distinction
between the other published models is infeasible because uncertainties in the
initial conditions, in the $N$-body simulations themselves (e.g., artificial
heating), and in the assumed model parameters (potential, secular heating, star
formation histories, etc.) can produce differences that are roughly comparable
to the predicted differences {\it between} various scenarios.

\section{Summary and Conclusions}

We have assembled a sample of $\sim$17,300 G-type dwarfs with available
low-resolution ($R\sim$2000) spectroscopy from SEGUE, a sub-survey
conducted during SDSS-II. The sample we considered comprises stars with $d < 3$
kpc, \logg\ $\geq 4.2$, spectra having S/N $\geq 30$ \AA$^{-1}$, \vphi\ $> +50$ \kms, [Fe/H]
$>$ $-$1.2, and $7 < R < 10$ kpc. A separate test was carried out to eliminate a
small number of stars that had larger probability of being associated with the
halo than the disk system.

Unlike the conventional assignment of stars into thin- and thick-disk
components based on kinematics (or spatial distribution), we have made use
of \afe\ as a reference to {\it chemically divide} our G-dwarf sample into likely
thin-disk and thick-disk populations.

Our chemically separated populations indicate that a negative rotational
velocity gradient with increasing [Fe/H] exists for the thin-disk population
($-22.6 \pm 1.6$ \kmsdex), while the thick-disk population exhibits a positive slope
($+45.8 \pm 2.9$ \kmsdex) in the range $0.1 < |Z| < 3.0$ kpc and $R$ = 7--10 kpc. Larger
mean orbital radii are also noticed among the metal-poor thin-disk stars, as
compared to the more metal-rich thin-disk stars, and smaller mean orbital radii
are found for the thick-disk stars compared with the thin-disk stars.

The distribution of rotational velocity appears independent of $R$ for our
thin-disk subsample, while there exists a very small correlation ($-5.6 \pm 1.1$ 
\kmskpc) between \vphi\ and $R$ for our thick-disk subsample.

We have found that the observed lag of \vphi\ for the high-\afe\ stars relative
to the low-\afe\ population is quite constant at a given \z\ distance (30 \kms),
implying that our chemically separated populations are indeed distinct
components in terms of their kinematics. This also allows us to infer that
division by chemistry reveals the kinematic structure of each population better
than division on the basis of spatial separation.

The vertical gradient of \vphi\ with \z\ for our thick-disk subsample ($-9.4 \pm 1.3$ 
\kmskpc) is smaller than that reported by Casetti-Dinescu et al. (2011; $-$25
\kmskpc), Ivezi\'c et al. (2008; $-$29 \kmskpc), Girard et al. (2006; $-30$
\kmskpc), and Chiba \& Beers (2000; $-$30 \kmskpc). Without application of our proposed
separation of the thin- and thick-disk subsamples, we find a vertical gradient
of $-19.4 \pm 1.8$ \kmskpc\ for the stars with \z\ $> 1.0$ kpc, in 
better agreement with the previous observational studies.
Hence, depending on how the thick-disk stars are selected (spatially or
chemically), the kinematic trends change. Assuming the thick disk is a distinct
component comprised of stars with high-\afe\ ratios, we again stress that the
chemical separation of the thick disk provides a more clear picture of the
kinematics.

It appears that there is no correlation between orbital eccentricity and metallicity
for the thin-disk subsample, while the trend of $e$ for the thick-disk subsample
rather steeply increases as the metallicity decreases. The $e$ distribution 
for the low-\afe\ stars appears to be
independent of $R$, whereas the high-\afe\ stars exhibit an increasing trend
with distance from the Galactic center. The difference in average orbital
eccentricity between the low-\afe\ and high-\afe\ subsamples appears constant
at any given \z\ (about 0.1), which also indicates a clear distinction between
these populations. Our approach of separating the thin-disk and thick-disk 
components by chemical tagging on their \afe\ abundance ratios yields 
well-defined, and distinct, kinematic trends for these populations as 
listed in Table \ref{tab:obs}. 

The rotational velocity gradient for the thin-disk subsample with metallicity
qualitatively agrees with the predictions of the radial migration models
(Sch\"onrich \& Binney 2009b; Loebman et al. 2010). Table \ref{tab:test}
summarizes the results of qualitative comparisons of the various thick-disk
formation scenarios with the observed properties of our
G-dwarf sample, based on predictions from these published (but still rather
primitive) models.

Based on these results, radial migration appears to have influenced the
structural and chemical evolution of the thin disk, but may not have played a
prominent role in the formation and evolution of the thick disk. However, to be
certain of this inference, comparisons with the predictions
of more refined radial migration models that better reproduce the observed
properties of the thick-disk population of the Milky Way galaxy are required.
The preponderance of evidence, based on qualitative comparisons with existing
thick-disk formation models, indicates that the thick disk of the Milky Way may
have resulted from gas-rich mergers, or from heating of a pre-existing thin disk
by minor mergers. We again emphasize that, although all of the models considered
have had some success in reproducing aspects of the thick disk, no one theory
has emerged that fully accounts for its detailed observed properties. We expect
that newer generation models and simulations will be strongly constrained by
observations such as those presented here.

\acknowledgments

Funding for SDSS-I and SDSS-II has been provided by the Alfred P.
Sloan Foundation, the Participating Institutions, the National Science
Foundation, the U.S. Department of Energy, the National Aeronautics
and Space Administration, the Japanese Monbukagakusho, the Max Planck
Society, and the Higher Education Funding Council for England.  The SDSS
Web Site is http://www.sdss.org/.

This work was supported in part by grants PHY 02-16783 and PHY 08-22648: Physics
Frontiers Center/Joint Institute for Nuclear Astrophysics (JINA), awarded by the
U.S. National Science Foundation. J.A.J. acknowledges support from NSF grant
AST-0607482. \v Z.I. acknowledges support from NSF grants AST 06-15991 and AST-07 07901, as
well as from grant AST 05-51161 to LSST for design and development activities.

%\clearpage

\appendix
\section{Resolving Observational Conflicts on
Correlations between Rotational Velocity and Metallicity}

The recent study of Ivezi\'c et al. (2008) concluded, on the basis of their
analysis of a complete photometric sample from SDSS, that there existed little
or no correlation between \vphi\ and [Fe/H] for stars in the disk system of the
Milky Way, a finding confirmed by Bond et al. (2010). However, based on a
spectroscopic sample of dwarfs from SDSS DR7, Spagna et al. (2010) reported a
gradient of 40--50 \kmsdex\ for stars with $-1.0 <$ [Fe/H] $< -0.5$ and $1<$ \z\
$< 3$ kpc. Loebman et al. (2010) claimed that the gradient found by Spagna et
al. (2010) was caused by selection bias in the SDSS spectroscopic sample. As our
current analysis also reveals a trend of \vphi\ with [Fe/H] for likely
thick-disk stars, of similar size to that reported by Spagna et al., we have
attempted to resolve these contradictory results between the various studies.

\begin{figure*}[b!]
\medskip
\centering
%\plotone{zeljko_vs_mine_diff_small.eps}
\includegraphics[height=120mm,width=160mm]{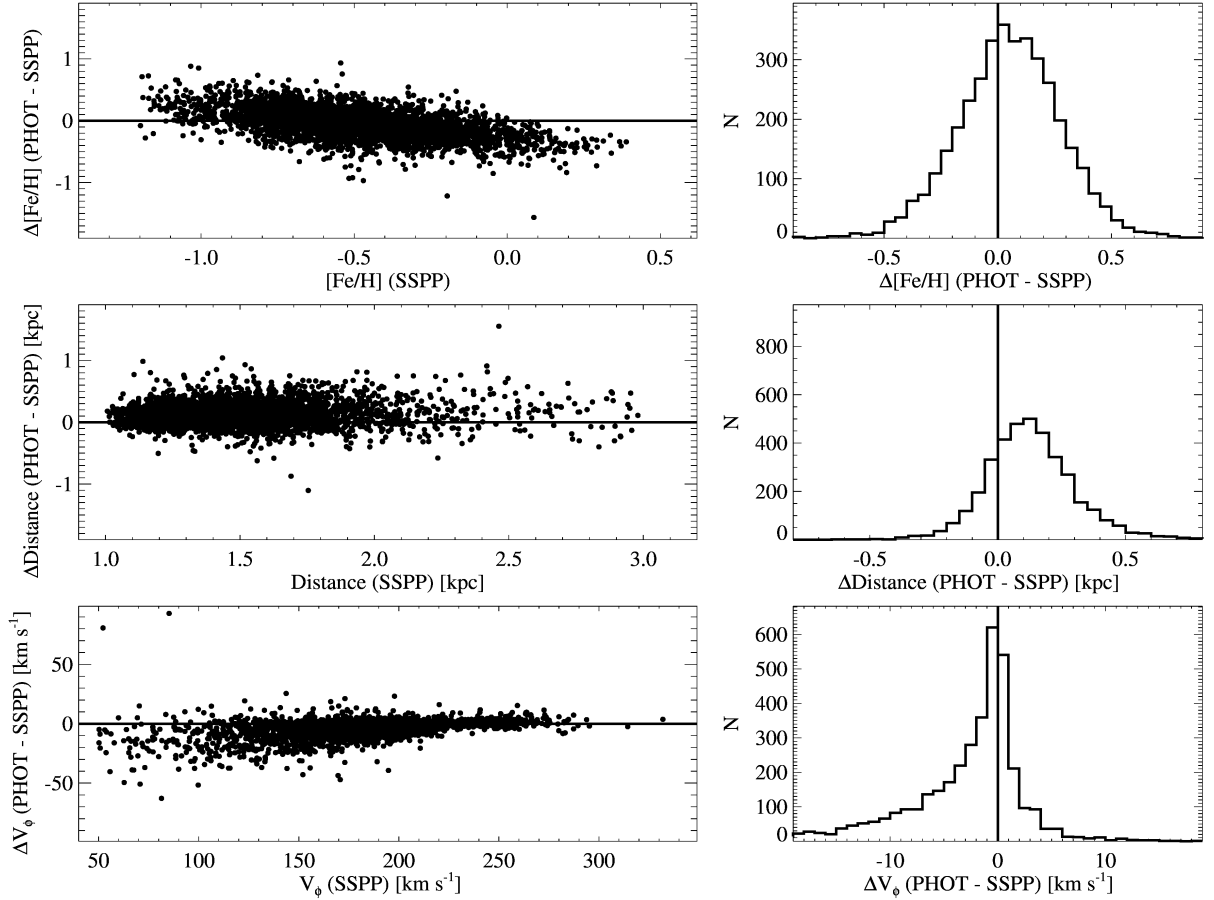}
\caption{Scatter plots (left panels) and histograms (right panels) of the
differences in \feh, distance, and \vphi, from top to bottom, between the
photometric estimates (PHOT) and our estimates (SSPP) (see text for details of
the determination of the ``PHOT'' parameters).}
\label{fig:zeljkodiff}
\end{figure*}
%\newpage

\begin{figure}[t!]
\centering
%\plotone{feh_vphi_zeljko_mine.eps}
%\includegraphics[scale=0.7]{feh_vphi_zeljko_mine.eps}
\includegraphics[height=150mm,width=120mm]{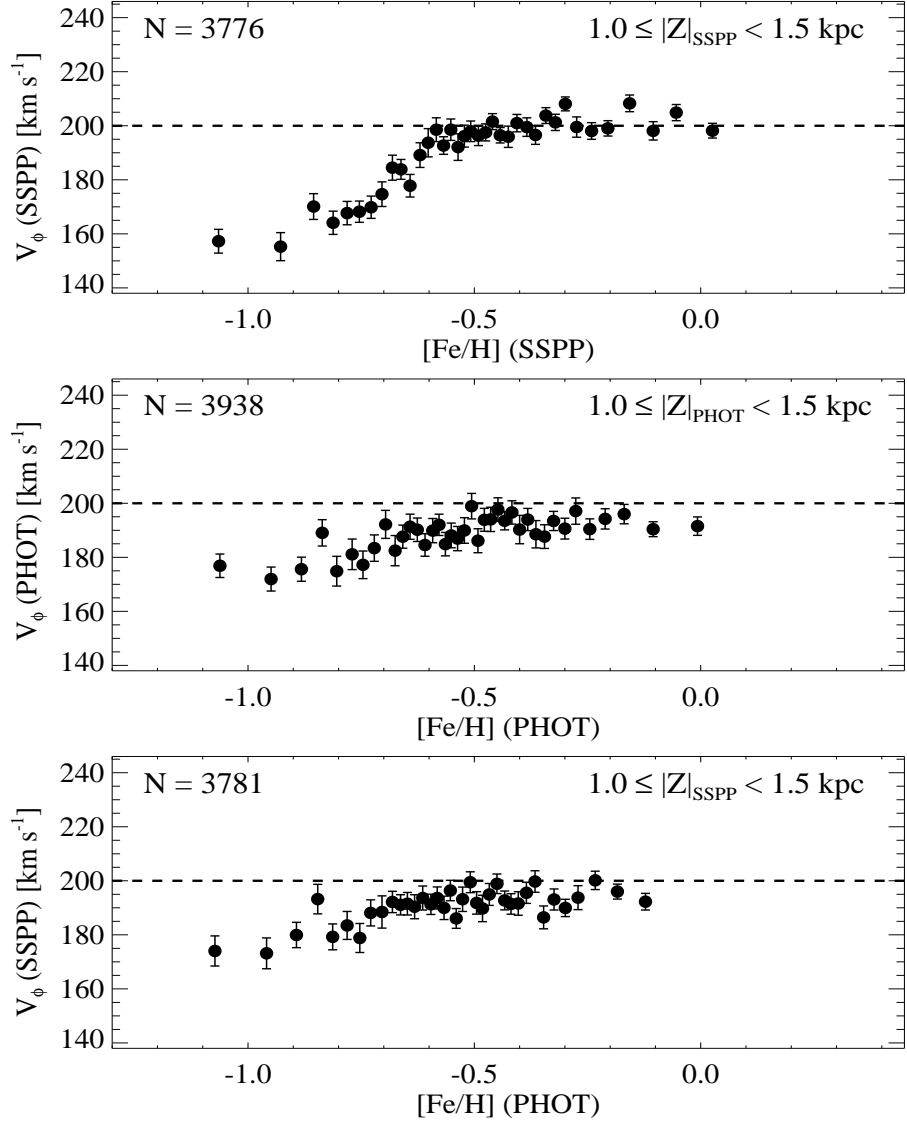}
\caption{Derived rotational velocity gradients as a function of metallicity.
The top panel shows the trend with the SSPP-derived quantities for all
stars at \z\ = 1.0--1.5 kpc from our G-dwarf
sample. Note that the SSPP-determined \z\ distance is used to select these
stars. The middle panel are the photometric determinations in the same
interval of height above the plane. Notice that there are more objects selected
in this \z\ distance range, which was calculated using Equation
\ref{eqn:dist}. The bottom panel displays the trend of our values of \vphi\ vs.
the photometric metallicity, and a very similar pattern between the middle and
the bottom panel is observed. The error bar on
each point is calculated by  resampling 100 stars with replacement 1000 times.}
\medskip
\label{fig:zeljko}
\end{figure}

As a first step, we employ the relationship devised by Ivezi\'c et al. (2008)
to obtain absolute magnitudes in the $r$ band, and derive distances for our
G-dwarf sample that should be on the same scale as theirs. Their adopted
relationship is:

\begin{equation}
\begin{split}
  M_{r}(g-i, [\rm Fe/H]) = & -0.56 + 14.32 \, x -12.97 \, x^{2} \\
                           & + 6.127 \, x^{3}-1.267 \, x^{4}+0.0967 \, x^{5} \\
                           & -1.11 \, [\rm Fe/H]-0.18 \, [\rm Fe/H]^{2},
\end{split}
\label{eqn:dist}
\end{equation}

\noindent where $x=(g-i)$. The above is the combined relationship of Equations A2, A3, and A7 in
Ivezi\'c et al. (2008). Then, by adopting the improved expression by
Bond et al. (2010), we estimate the photometric metallicities for our sample.
The adopted relationship is as follows:

\begin{equation}
\begin{split}
  [\rm Fe/H]_{PHOT} = & -13.13+14.09 \, x+28.04 \, y-5.51 \, xy \\
                      & -5.90 \, x^{2}-58.68 \, y^{2}+9.14 \,x^{2}y \\
                      & -20.61 \, xy^{2}+0.00 \, x^{3}+ 58.20 \, y^{3},
\end{split}
\label{eqn:feh}
\end{equation}

\noindent where $x=(u-g)$ and $y=(g-r)$. All colors are reddening corrected,
and note that the coefficient of the $x^{3}$ term is zero. We refer to the
distance determined with Equation \ref{eqn:dist} as the ``photometric
distance'', the metallicity estimated by Equation \ref{eqn:feh} as the
``photometric metallicity'', and the rotation velocity calculated using the
photometric distance in combination with the measured radial velocities and
proper motions as the ``photometric rotational velocity''. The label ``PHOT'' in Figures
\ref{fig:zeljkodiff} and \ref{fig:zeljko} indicate these estimates, while
the label ``SSPP'' denotes the values we have used for the G-dwarf sample.

%\newpage

Figure \ref{fig:zeljkodiff} shows scatter plots (left panels) and histograms
(right panels) of the differences in \feh, distance, and \vphi\ between the
photometric (PHOT) estimates and our estimates (SSPP) for stars in \z\ = 1.0--1.5 kpc
from our G-dwarf sample considered in this paper. It can be noticed from inspection of the top panels
that the photometric metallicity is consistently higher at low [Fe/H] and lower
at high [Fe/H], compared to the SSPP estimates, with an overall shift of about
0.1 dex (top right panel).

The middle panels suggest that our distance
determination is on average $\sim$0.1 kpc lower than the photometric estimate,
with a trend that this deviation becomes larger as the distance increases. The
bottom panels show that our rotational velocities generally agree with the
photometric rotational velocities, but our values trend higher at low \vphi.

With these differences (especially for [Fe/H]) kept in mind, we have examined
the trend of \vphi\ with \feh\ based on the photometric estimates and based on
the spectroscopic estimates, for stars in the range \z\ = 1.0--1.5 kpc. Figure
\ref{fig:zeljko} shows the results. The top panel makes use of the
spectroscopically-derived quantities, while the middle panel comes from the
photometric quantities. Due to the small difference in the distance estimates
(larger distances for the photometric distance), we see
in the middle panel that there are more objects selected in this \z\ distance
range, which does not affect our conclusions. It is obvious that we obtain a
flattening of the \vphi\ relationship below [Fe/H] $< -0.5$ for the
photometrically-determined values. Even if we consider the photometric
metallicity and {\it our values} of \vphi, we obtain a very similar pattern
(bottom panel). This strongly suggests that the effect of the input metallicity
in Equation \ref{eqn:dist} on distance for calculation of \vphi\ is minimal.
However, it is clear that the difference between use of the photometric
metallicity and the spectroscopic metallicity makes a large difference in the
derived trend of \vphi. This can be accounted for by the scattering of higher
metallicity stars (which have high \vphi) into the photometrically determined
low-metallicity region, resulting in a flatter gradient of \vphi\ with \feh\ for
stars with \feh\ $< -0.5$, as seen by comparing the middle panel of Figure
\ref{fig:zeljko} with the top panel.

There are at least two reasons that the photometric metallicity
relation in Equation \ref{eqn:feh} may assign the metallicity of a star to a value
that strongly deviates from that estimated from the SSPP. First, Equation
\ref{eqn:feh} was not calibrated with the metallicity estimates used in this
study, but with those available in DR7. There has been significant improvement
in the SSPP for estimation of [Fe/H] since the DR7 release, especially for stars
with near-solar and super-solar values (see Smolinski et al. 2011). For a proper
comparison, the photometric metallicity relation needs to be re-calibrated with
the metallicities available from the DR8 release. The other reason
is that small random errors in the photometric measurements (and zero points) can
strongly influence the photometric metallicity estimate 
(and its errors) (\v Z. Ivezi\'c et al. in preparation). As a result, stars 
that are in reality of high metallicity can be artificially
moved into the low-metallicity region.

\begin{thebibliography}{}
\bibitem[]{} Abadi, M. G., Navarro, J. F., Steinmetz, M. \& Eke, V. R. 2003, \apj, 597, 21
\bibitem[]{} Abazajian, K., et al. 2009, \apjs, 182, 543
\bibitem[]{} Andrievsky, S. M., Luck, R. E., Martin, P., \& L\'epine, J. R. D. 2004, \aap, 413, 159
\bibitem[]{} Aihara, H., et al. 2011, \apjs, 193, 29
\bibitem[]{} Allende Prieto, C., et al. 2008, \aj, 136, 2070
\bibitem[]{} An, D., Terndrup, D. M., Pinsonneault, M. H., Paulson, D. B.,
              Hanson, R. B., \& Stauffer, J. R. 2007, \apj, 655, 233
\bibitem[]{} An, D., et al. 2009a, \apj, 700, 523
\bibitem[]{} An, D., et al. 2009b, \apjl, 707, L64
\bibitem[]{} Barbanis, B., \& Woltjer, L. 1967, \apj, 150, 461
\bibitem[]{} Bensby, T., Feltzing, S., \& Lundstr\"om, I. 2003, \aap, 410, 527
\bibitem[]{} Bensby, T., Feltzing, S., Lundstr\"om, I., \& Ilyin, I. 2005, \aap, 433, 185
\bibitem[]{} Bond, N. A., et al. 2010, \apj, 716, 1
\bibitem[]{} Brook, C. B., Kawata, D., Gibson, B. K., \& Freeman, K. C. 2004, \apj, 612, 894
\bibitem[]{} Brook, C. B., Gibson, B. K., Martel, H. \& Kawata, D. 2005, \apj, 630, 298
\bibitem[]{} Brook, C., Richard, S., Kawata, D., Martel, H., \& Gibson, B. K. 2007, \apj, 658, 60
\bibitem[]{} Carollo, D., et al. 2010, \apj, 712, 692
\bibitem[]{} Casetti-Dinescu, D. I., Girard, T. M., Korchagin, V. I., \& van Altena, W. F.
             2011, \apj, 728, 7
\bibitem[]{} Chiba, M., \& Beers, T. C. 2000, \aj, 119, 2843
\bibitem[]{} Di Matteo, P., Lehnert, M. D., Qu, Y., \& van Driel, W. 2011, \aap, 525, L3
\bibitem[]{} Dierickx, M., Klement, R. J., Rix, H.-W., \& Liu, C. 2010, \apj, 725, L186
\bibitem[]{} Fuchs, B. 2001, \mnras, 325, 1637
\bibitem[]{} Fuhrmann, K. 1998, \aap, 338, 161
\bibitem[]{} Fuhrmann, K. 2008, \mnras, 384, 173
\bibitem[]{} Gilmore, G., \& Reid, N. 1983, \mnras, 202, 1025
\bibitem[]{} Gilmore, G., Wyse, R. F. G., \& Bryn Jones, J. 1995, \aj, 109, 3
\bibitem[]{} Girard, T. M., Korchagin, V. I., Casetti-Dinescu, D. I., van Altena, W. F.,
             L\'opez, C. E., \& Monet, D. G. 2006, \aj, 132, 1768
\bibitem[]{} Girard, T. M., et al. 2011, \aj, 142, 15
\bibitem[]{} Haywood, M. 2001, \mnras, 325, 1365
\bibitem[]{} Haywood, M. 2008, \mnras, 388, 1175
\bibitem[]{} Holmberg, J., Nordstr\"om, B., \& Andersen, J. 2007, \aap, 475, 519
\bibitem[]{} Ivezi\'c, \v Z. et al. 2008, \apj, 684, 287
\bibitem[]{} Jenkins, A. 1992, \mnras, 257, 620
\bibitem[]{} Kazantzidis, S., Bullock, J. S., Zentner, A. R., Kravtsov, A. V., \&
             Moustakas, L. A. 2008, \apj, 688, 254
\bibitem[]{} Lemasle, B., Francois, P., Bono, G., Mottini, M., Primas, F., \& Romaniello, M. 2007, \aap, 467, 283
\bibitem[]{} Lee, Y. S., et al. 2008a, \aj, 136, 2022
\bibitem[]{} Lee, Y. S., et al. 2008b, \aj, 136, 2050
\bibitem[]{} Lee, Y. S., et al. 2011, \aj, 141, 90
\bibitem[]{} Loebman, S. R., Ro\v skar, R., Debattista, V. P., Ivezi\'c, \v Z., Quinn, T., \&
             Wadsley, J. 2010, ArXiv:1009.5997
\bibitem[]{} Magrini, L., Sestito, P., Randich, S., $\&$ Galli, D. 2009, \aap, 494, 95
\bibitem[]{} Majewski, S. R. 1993, \araa, 31, 575
\bibitem[]{} Minchev, I., \& Famaey, B. 2010, \apj, 722, 112
\bibitem[]{} Munn, J., et al. 2004, \aj, 127, 3034
\bibitem[]{} Munn, J., et al. 2008, \aj, 136, 895
\bibitem[]{} Navarro, J. F., Abadi, M. G., Venn, K. A., \& Freeman, K. C. 2011, \mnras, 412, 1203
\bibitem[]{} Nissen, P. E., \& Schuster, W. J. 2010, \aap, 511, L10
\bibitem[]{} Nordstr\"om, B., et al. 2004, \aap, 418, 989
\bibitem[]{} Parker, J. E., Humphreys, R. M., \& Beers, T. C. 2004, \aj, 403, 74
\bibitem[]{} Quinn, P. J., Hernquist, L., \& Fullagar, D. P. 1993, \apj, 403, 74
\bibitem[]{} Reddy, B. E., Lambert, D. L., \& Allende Prieto, C. 2006, \mnras, 367, 1329
\bibitem[]{} Reddy, B. E. 2010, in IAU Symp. 265, Chemical Abundances in the Universe:
             Connecting First Stars to Planets, ed.
             K. Cunha, M. Spite, \& B. Barbuy (Cambridge: Cambridge Univ. Press), 289
\bibitem[]{} Richard, S., Brook, C. B., Martel, H., Kawata, D., Gibson, B. K.,
             Sanchez-Blazquez, P. 2010, \mnras, 402, 1489
\bibitem[]{} Robin, A. C., Reyl\'e, C., Derri\'ere, S., \& Picaud, S. 2003, \aap, 409, 523
\bibitem[]{} Rocha-Pinto, H. J., Rangel, R. H. O., Porto de Mello, G. F., Braganca, G. A., \& Maciel W. J. 2006, 453, L9
\bibitem[]{} Ro\v skar, R., Debattista, V. P., Stinson, G. S., Quinn, T. R., Kaufmann, T., \& Wadsley, J. 2008a, 675, L65
\bibitem[]{} Ro\v skar, R., Debattista, V. P., Quinn, T. R., Stinson, G. S., \& Wadsley, J. 2008b, \apj, 684, L79
\bibitem[]{} Sales, L. V., et al. 2009, \mnras, 400, L61
\bibitem[]{} Schlesinger, K. J., Johnson, J. A., Lee, Y. S., Masseron, T., Yanny, B., Rockosi, C. M.,
             Gaudi, B. S., \& Beers, T. C. 2010, \apj, 719, 996
\bibitem[]{} Sch\"onrich, R., \& Binney, J. 2009a, \mnras, 396, 203
\bibitem[]{} Sch\"onrich, R., \& Binney, J. 2009b, \mnras, 399, 1145
\bibitem[]{} Sch\"onrich, R., Binney, J., \& Dehnen, W. 2010, \mnras, 403, 1829
\bibitem[]{} Sellwood, J. A., \& Binney, J. J. 2002, \mnras, 336, 785
\bibitem[]{} Sesar, B., Ivezi\'c, \v Z., \& Juri\'c, M. 2008, \apj, 689, 1244
\bibitem[]{} Sestito, P., Bragaglia, A., Randich, S., Pallavicini, R., Andrievsky, S. M., \& Korotin, S. A. 2008, \aap, 488, 943
\bibitem[]{} Steinmetz, M., et al. 2006, \aj, 132, 1645
\bibitem[]{} Soubiran, C., Bienaym\'e, O., \& Siebert, A. 2003, \aap, 398, 141
\bibitem[]{} Soubiran, C., \& Girard, P. 2005, \aap, 438, 139
\bibitem[]{} Spagna, A., Lattanzi, M. G., Re Fiorentin, P., \& Smart, R. L. 2010, \aap, 510, L4
\bibitem[]{} Spitzer, L., \& Schwarzschild, M. 1953, \apj, 118, 106
\bibitem[]{} Smolinski, J. P., et al. 2011, \aj, 141, 89
\bibitem[]{} Villalobos, \'A., \& Helmi, A. 2008, \mnras, 391, 1806
\bibitem[]{} Villalobos, \'A., Kazantzidis, S., \& Helmi, A. 2010, \apj, 718, 314
\bibitem[]{} Wilson, M., et al. 2011, \mnras, 413, 2235
\bibitem[]{} Wyse, R. F. G., Gilmore, G., Norris, J. E., Wilkinson, M. I., Kleyna, J. T.,
             Koch, A., Evans, N. W., \& Grebel, E. K. 2006, \apj, 639, L13
\bibitem[]{} Yanny, B., et al. 2009, \aj, 137, 4377
\bibitem[]{} York, D. G., et al. 2000, \aj, 120, 1579
\bibitem[]{} Yoshii, Y. 1982, \pasj, 34, 365
\bibitem[]{} Zwitter, T., et al. 2008, \aj, 136, 421
\end{thebibliography}
\end{document}